\documentclass[journal=jpcafh,manuscript=article]{achemso}
\usepackage{chemformula} 
\usepackage[T1]{fontenc} 
\usepackage{amsmath}
\usepackage{graphicx}
\usepackage{array}
\usepackage[normalem]{ulem}
\usepackage[table-number-alignment=center]{siunitx}
\usepackage{multirow}
\usepackage{setspace}

\newcolumntype{M}[1]{>{\centering\arraybackslash}m{#1}}
\newcommand{\onlinecite}[1]{\hspace{-1 ex} \nocite{#1}\citenum{#1}}

\author{Piotr Michalak}
\author{Michał Lesiuk}
\affiliation[University of Warsaw]
{University of Warsaw, Faculty of Chemistry, Pasteura 1, Warsaw, 02-093, Poland}
\email{m.lesiuk@uw.edu.pl}

\title{Rank-reduced equation-of-motion coupled cluster formalism with full inclusion of triple excitations}

\begin{document}

\begin{tocentry}
 \includegraphics[width=8.0cm, height=4.4cm]{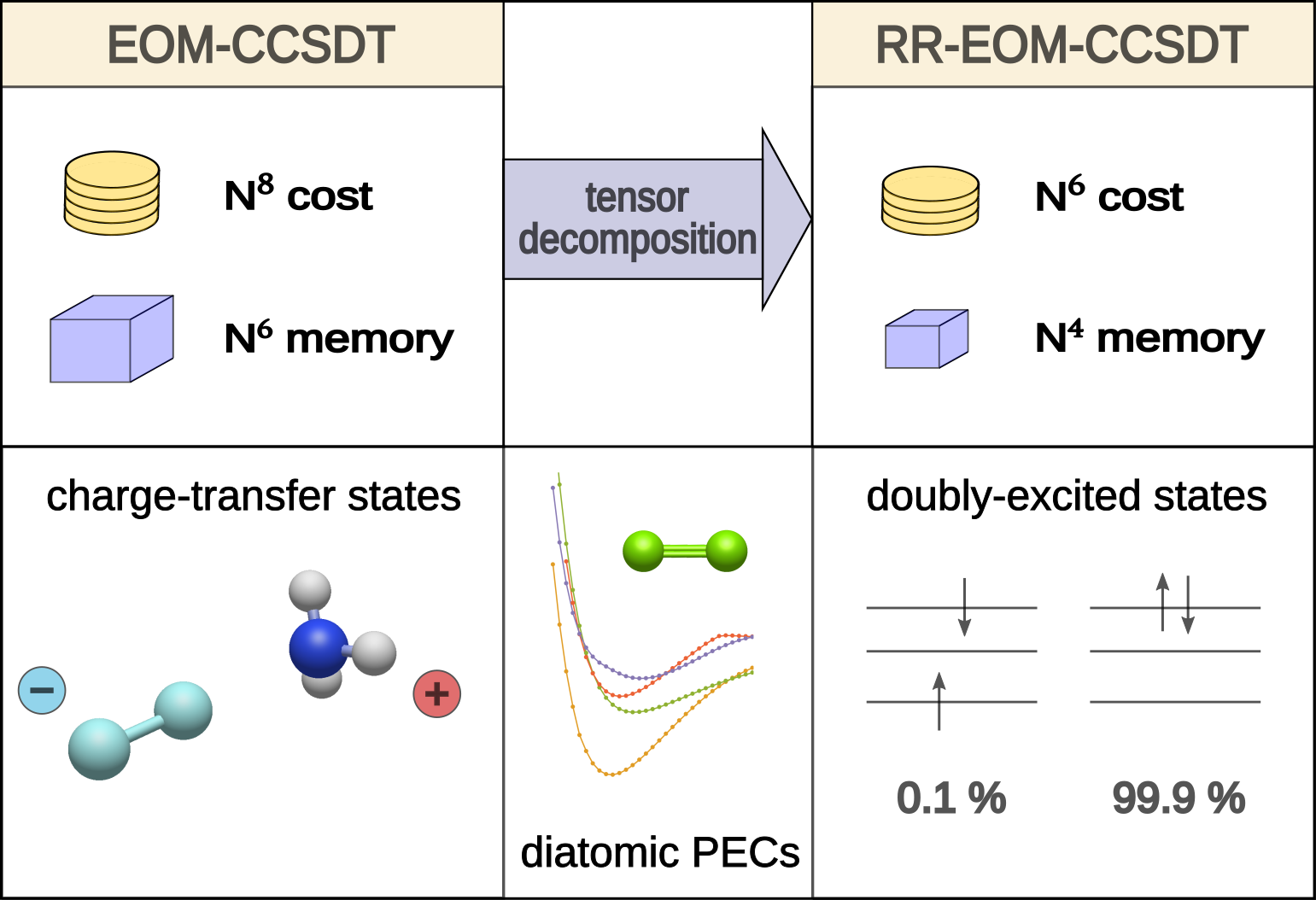}
\end{tocentry}

\begin{abstract}
In this work we describe the rank-reduced variant of the equation-of-motion coupled cluster theory with complete inclusion of single, double, and triple excitations. The advantage of the proposed formalism in comparison with the canonical theory stems from the application of Tucker decomposition format to the ground- and excited-states triply-excited amplitudes tensors. By exploiting the linear scaling of the dimension of the decomposed amplitudes with respect to the system size $N$, one can reduce the computational cost of the method to the level of $N^6$ and storage requirements to $N^4$. While in practice the proposed rank-reduced formalism introduces an error, we show that it is several times smaller than the inherent error of the parent theory with the proposed default settings for a wide range of problems. Higher level of accuracy can be achieved by increasing the value of a single parameter present in this formulation, recovering the canonical method in an appropriate limit. We illustrate the accuracy and performance of the proposed method by calculations for a group of molecules with excited states of different character -- from dominated by single excitations with respect to the reference determinant to states with moderate and large contribution of double and higher excitations. We report calculations of potential energy curves and related spectroscopic parameters for the first four singlet excited states of magnesium dimer, as well as the potential energy curve for the excited state of charge-transfer character in the NH$_3$-F$_2$ complex as a function of intermolecular separation.
\end{abstract}
\newpage

\section{Introduction}
\label{sec:intro}
Molecular excited states lie at the heart of many research areas, such as design of new materials for photovoltaics or photocatalysis, that are of both academic and industrial interest. It is therefore not surprising that significant effort has been focused on development of reliable and affordable theoretical methods directed at description of excited states and related chemical transformations. Quantum-chemical approaches that target excited states are numerous and are constantly being proposed and improved; for an in-depth overview of this topic we refer to the recent review articles~\cite{serrano2005,dreuw2005,gonzalez2012,lischka2018,mai2020}. Broadly speaking, the status of the most frequently used method of this class is firmly occupied by the time-dependent density functional theory~\cite{casida1995,dreuw2005,casida2012} (TDDFT) due to its comparatively low cost and conceptual simplicity, among other factors. However, TDDFT is often criticized for the lack of systematic improvability, large sensitivity to the choice of the exchange-correlation functional approximation, and need for large-scale benchmarking and comparison with external reference data~\cite{laurent2013,bremond2018,loos2019,sarkar2021} to arrive at meaningful results. An alternative class of methods which is, at least in principle, free from these problems is based on the coupled cluster (CC) ansatz of the electronic wave function~\cite{crawford00,bartlett07}.

The CC theory has been successfully applied to excited states either with help of the equation-of-motion CC (EOM-CC) formalism~\cite{stanton93,krylov08} or using the linear response approach~\cite{monkhorst77,dalgaard83,sekino84,koch90,koch94}. As long as only the excitation energies are considered, as is throughout the present work, these two approaches are fully equivalent. EOM-CC results can be systematically improved towards the exact solution of the electronic Schr\"odinger equation for any state by including higher and higher excitations with respect to the reference determinant. However, this comes at a steep computational cost. For example, inclusion of all single and double excitations leads to the EOM-CCSD method~\cite{stanton93} which scales as $N^6$ with the system size, $N$. While the accuracy of this method may be acceptable in applications when excited states dominated by single excitations are considered, even for such states inclusion of triply-excited configurations is necessary for benchmark-quality results. Moreover, it is known that EOM-CCSD struggles when post-excitation orbital relaxation effects are large~\cite{subotnik11}, e.g. in charge-transfer states, or for states dominated by double excitations with respect to the reference determinant. In such problematic cases, unacceptable errors of the order of several eV are not uncommon in the EOM-CCSD results and inclusion of triple excitations, or even higher, becomes essential. Unfortunately, inclusion of triple excitations increases the computational scaling to $N^7$ when they are treated approximately, e.g. iteratively as in CCSDT-$n$~\cite{lee1984,urban1985,noga1987}, CC3~\cite{christiansen1995,koch1997} or $n$CC methods~\cite{bartlett2006,teke2024} or non-iteratively as an additive correction~\cite{watts1995,watts1996,stanton1996,christiansen1996,hirata2001,kowalski2004,manohar2009,sauer2009,watson2013,matthews16}, while the complete EOM-CCSDT theory~\cite{noga87,scuseria88,kowalski01,kucharski01} scales as $N^8$. In any case, this makes application of methods that include triple excitations to systems with more than a dozen or so non-hydrogen atoms a daunting proposition (assuming a decent-quality basis set is used).

In this paper, we follow-up on our previous work devoted to the EOM-CC3 method~\cite{michalak24} and demonstrate how to reduce the cost of the EOM-CCSDT calculations by using tensor decomposition techniques~\cite{kolda09,lathauwer2000} applied to the triply-excited amplitudes. While the idea of reducing the rank of CC amplitudes using various decomposition formats is actively pursued by several groups, see Refs.~\onlinecite{kinoshita03,hino04,scuseria08,bell10,hohenstein12,parrish12,parrish13a,parrish13b,mayhall17,schutski17,lesiuk2019,hohenstein19,parrish19,Lesiuk2020,hohenstein2022,Lesiuk2022,lesiuk2022b} and references therein, the attention has been focused mostly on the ground-state calculations thus far and applications to molecular excited states are much more limited~\cite{hohenstein13,hohenstein19,michalak24}. Here, we report the rank-reduced variant of the EOM-CCSDT theory with scaling proportional to $N^6$ with respect to the system size and hence applicable to larger systems at a decreased cost in comparison with the canonical counterpart. There are several major reasons that motivate this development. First, when high-accuracy results are sought after, EOM-CCSDT method is frequently used as an ingredient in composite schemes that account for higher-order correlation effects additively (see the papers of Loos and collaborators~\cite{loos19,loos23} for excellent examples of this methodology). Second, the rank-reduced EOM-CCSDT method forms a basis for further extensions towards, for example, EOM-CC4 theory~\cite{kallay04,kallay05} that has been recently singled out as a method capable of delivering near-FCI quality results for a considerable range of excited state types~\cite{loos21,loos22}. Finally, the experience gathered here is likely to be transferable to other excited-state methods that include triple excitations in the wave function ansatz, such as high-order adiabatic diagrammatic construction (ADC)~\cite{schirmer1982,dreuw2015,herbst2020,dreuw2023}, renormalized~\cite{kowalski02,piecuch09,piecuch15} and stochastic CC approaches~\cite{shen12a,shen12b,deustua19,yuwono20}, or distinguishable clusters models~\cite{kats2013,kats2014,rishi2019}, with comparable reductions of their computational costs.

\section{Theory}
\label{sec:theory}

\subsection{Notation and basics}

In this section we introduce the notation employed in this work and recall the basic formalism of the equation-of-motion coupled cluster theory. The convention used throughout the text for the indices appearing in the equations is summarized in Table~\ref{notation}. Additionally, we use the Einstein convention (implicit summation over repeating indices) in the present work unless explicitly stated otherwise. 

In this paper we are only concerned with the electronic singlet states and hence the starting point of our developments is the closed-shell, canonical Hartree-Fock determinant, denoted $|0\rangle$. The singly-, doubly-, triply-, etc., excited determinants are denoted by $|_i^a\rangle$, $|_{ij}^{ab}\rangle$, $|_{ijk}^{abc}\rangle$, and so on, and the orbital energies are denoted by $\epsilon_p$. We use the normal-ordered form of the electronic Hamiltonian, $H_N = H - E_{HF}$, with respect to $|0\rangle$, where $E_{HF}$ is the Hartree-Fock energy. The customary partitioning of this operator is adopted, $H_N = F_N + V_N$, where we have introduced the normal-ordered Fock operator $F_N$ and fluctuation potential $V_N$. We take advantage of the $T_1$-similarity-transformed formalism, where the transformation is denoted with a tilde and defined for an arbitrary operator $A$ as $\tilde{A}$ = $e^{-T_1} A e^{T_1}$. In particular, the $T_1$-transformed electron repulsion integrals (ERIs) in the Coulomb notation are denoted by $(pq\widetilde{|}rs)$. In our implementation we use the density-fitting approximation\cite{katouda2009} which allows us to decompose ERIs according to the following formula:
\begin{align}
(pq\widetilde{|}rs) = B_{pq}^Q \,B_{rs}^Q, 
\end{align}
where the tensors $B_{pq}^Q$ are defined through the three-center, $(pq\widetilde|P)$, and two-center, $V_{PQ}$, electron repulsion integrals as:
\begin{align} \label{BpqQ}
 B_{pq}^Q = ( pq \widetilde{|} P )\, [\mathbf{V}^{-\frac{1}{2}}]_{PQ}.   
\end{align}
For brevity, in the present work we use two permutation operators which allow us to write the equations in a more compact form. The first is the basic permutation operator $P_{ij}^{ab}$ that exchanges pairs of indices $ia$ and $jb$. The second is the operator $P_3$ defined as:
\begin{align}
\label{p3}
        P_3 &= \left(1 + P_{ij}^{ab} \right) \left(1 + P_{ik}^{ac} + P_{jk}^{bc}  \right),
\end{align}
which symmetrizes any expression it operates on with respect to all possible permutations of the pairs of indices $ia$, $jb$, and $kc$.

\begin{table}[h] 
\centering
\small
    \begin{tabular}{M{2cm} p{11.5cm} M{2cm}}
    \hline
          indices & \centering meaning  & range \\
          \hline
        $i,j,k,l, \ldots$  & active orbitals occupied in the reference  &  $O$ \\
        $a, b, c, d, \ldots$  & orbitals unoccupied in the reference (virtual)  &  $V$ \\
        $p, q, r, s, \ldots$  & general orbitals (occupation not specified)  &  $N$ \\
        $P, Q, \ldots$  & density fitting auxiliary basis set  &  $N_{\mathrm{aux}}$ \\
        $x, y, z, \ldots$  & compressed subspace of the triply-excited ground-state amplitudes  &  $N_{\mathrm{svd}}$ \\
        $X, Y, Z, \ldots$  & compressed subspace of the triply-excited excited-state amplitudes  &  $N_{\mathrm{SVD}}$ 
    \end{tabular}
    \caption{Details of the notation used in the text.}
    \label{notation}
\end{table}

\subsubsection{Equation-of-motion coupled cluster theory}

In the equation-of-motion coupled cluster theory~\cite{stanton93,krylov08} one assumes the following ansatz for the electronic wave function of a molecular excited state:
\begin{align}
\label{eomwfn}
\Psi = R\,e^T |0\rangle,
\end{align}
where $R$ and $T$ are the usual cluster operators and $|0\rangle$ denotes the reference determinant. The operators $R$ and $T$ are defined as the sum of $m$-tuple excitation operators $R_m$ and $T_m$:
\begin{align}
    R &= R_0 + R_1 + R_2 + \ldots + R_M, \\
    T &= T_1 + T_2 + \ldots + T_M,
\end{align}
where the sum formally extends to the number of electrons in the system, $M$. The excitation operators are, in turn, defined in the second quantization through the standard creation and annihilation operators as:
\begin{equation}
R_m = \frac{1}{(m!)^2}\ \sum_{\substack{a_1, a_2, ... , a_m \\i_1, i_2, ... , i_m }}R_{\,i_1, i_2, ... , i_m}^{\,a_1, a_2, ... , a_m} \,a_1^\dag\,a_2^\dag\ldots a_m^\dag i_m \ldots i_2\,i_1,
\end{equation}
and similarly for $T_m$. The only exception from the above definition is the $R_0$ term which is a state-dependent number. The EOM-CC equations are derived using the electronic Schr\"odinger equation upon inserting the ansatz~(\ref{eomwfn}) and projecting the resulting equation onto the reference wave function and sets of all singly-, doubly-, triply-, etc., excited determinants, giving:
\begin{align}
\label{eomeqs}
\begin{split}
    \langle 0|[\bar{H}_N, R]|0\rangle &= \omega\, \langle 0 | R |0\rangle, \\ 
    \langle _i^a|[\bar{H}_N, R]|0\rangle &= \omega\, \langle _i^a | R |0\rangle, \\
    \langle _{ij}^{ab}|[\bar{H}_N, R]|0\rangle &= \omega\, \langle _{ij}^{ab} | R |0\rangle, \\
    \langle _{ijk}^{abc}|[\bar{H}_N, R]|0\rangle &= \omega\, \langle _{ijk}^{abc} | R |0\rangle,
\end{split}
\end{align}
and so on. In the above we have introduced the similarity-transformed Hamiltonian, $\bar{H}_N = e^{-T} H_N e^T$, and denoted the excitation energy by $\omega$. This set of equations is equivalent to an eigenvalue problem $\mathcal{H} \mathcal{R} = \omega \mathcal{R}$, where $\mathcal{H}$ is the effective Hamiltonian matrix with elements defined block-wise as $\mathcal{H}_{mn}= \langle \mu_m | [\bar{H}_N, \mu_n]|0\rangle$. In this notation $\mu_n$ collectively denotes all $n$-tuple excitation operators and the vectors $\mathcal{R}$ consist of blocks gathering all singly-, doubly-, triply-, etc., excited amplitudes.

\subsection{EOM-CCSDT method}

The EOM-CCSDT method can be derived from the general EOM-CC theory introduced above by truncating the cluster operators $R$ and $T$ at the level of $R_3$ and $T_3$ excitation operators, respectively. After performing this truncation and expanding the $\bar{H}_N$ operator in the Baker-Campbell-Hausdorff (BCH) series, we obtain the EOM-CCSDT equations which read:

\begin{align}
    &\langle _i^a| \left( O_1^{\mathrm{CCSD}} + [\tilde{H}_N, R_3]
    \right)|0\rangle = \omega\, R_i^a, \label{singleeq}\\ 
    &\langle _{ij}^{ab}| \left(O_2^{\mathrm{CCSD}} + [\tilde{H}_N, R_3] + [[\tilde{H}_N, T_3], R_1]\right)|0\rangle = 
    \omega\, R_{ij}^{ab}, \label{doubleeq}  \\ \nonumber
    &\langle _{ijk}^{abc} | \Big( [[\tilde{V}_N, T_2] , R_1]    + [[\tilde{F}_N , T_3], R_1] + [[\tilde{V}_N, T_3] , R_1] + \frac{1}{2} [[[\tilde{V}_N, T_2], T_2] , R_1] + \nonumber \\ &+ [\tilde{V}_N , R_2] + [[\tilde{F}_N , T_2], R_2] + [[\tilde{V}_N, T_2] , R_2] + [[\tilde{V}_N, T_3] , R_2] + 
   [\tilde{F}_N , R_3] + \nonumber  \\ 
   &+ [\tilde{V}_N , R_3] + [[\tilde{V}_N, T_2] , R_3]  \Big) | 0 \rangle = \omega\, R_{ijk}^{abc}. \label{tripleeq}
\end{align}
In the first two equations, we write $O_1^{\mathrm{CCSD}}$ and $O_2^{\mathrm{CCSD}}$ to collectively denote all terms that appear already in the EOM-CCSD method. Explicit formulas for these terms can be found, for example, in Refs.~\onlinecite{stanton93,musiał2020}. The implementation of these terms in the proposed rank-reduced formalism is identical as in the standard EOM-CCSD theory, because the amplitudes appearing in the $T_1$, $T_2$, $R_1$, and $R_2$ operators are not approximated in any way in the CC/EOM-CC iterations. For a detailed analysis of the remaining terms in the first two above equations, i.e. $\langle _i^a|[\tilde{H}_N, R_3]|0 \rangle$, $\langle _{ij}^{ab}| [\tilde{H}_N, R_3] | 0 \rangle $ and $ \langle _{ij}^{ab} | [[\tilde{H}_N, T_3], R_1] |0 \rangle$, as well as their implementation in the rank-reduced context we refer to the previous paper~\cite{michalak24} on the RR-EOM-CC3 method. In the current work, these terms are treated in exactly the same way as in Ref.~\onlinecite{michalak24}.

In the present work we focus solely on the triple amplitudes equation (Eq.~\ref{tripleeq}) which contains additional terms in comparison with the EOM-CC3 method. In fact, these extra terms are responsible for the $N^8$ scaling of the computational costs of the EOM-CCSDT theory with the system size, $N$. Specifically, there is a large group of terms with either $O^5V^3$ or $O^4V^4$ scaling and a single term which scales as $O^3V^5$ (see Supporting Information). Note that the explicit formula for the triple amplitudes equation in the EOM-CCSDT method has already been described in the literature\cite{kucharski01,musiał2020}. However, there are typically numerous equivalent ways of rearranging the final equations and formulations provided by other authors may be difficult to trace here. To eliminate this inconvenience, in Supporting Information we include our working formulas of the EOM-CCSDT method which provide an opportune starting point for subsequent manipulations within the rank-reduced formalism.  

\subsection{Rank-reduced EOM-CCSDT}

\subsubsection{Overview}

The RR-EOM-CCSDT method proposed in this work relies on the Tucker decomposition\cite{tucker66,kolda09} of the ground-state and an excited-state triply-excited amplitudes tensors. The precise form of this decomposition reads
\begin{align}
    T_{ijk}^{abc} &= t_{xyz}\,U_{ia}^x\,U_{jb}^y\,U_{kc}^z,  \label{Tucker1} \\
    R_{ijk}^{abc} &= r_{XYZ}\,V_{ia}^X\,V_{jb}^Y\,V_{kc}^Z, \label{Tucker2}
\end{align}
where $t_{xyz}$ and $r_{XYZ}$ denote compressed amplitudes for the ground and an excited state, respectively, and the tensors $U_{ia}^x$ and $V_{ia}^X$ span the corresponding ground-state and excited-state triple-excitation subspaces. The sizes of these subspaces (lengths of summation over $x, y, z$ or $X, Y, Z$) are denoted as $N_{\mathrm{svd}}$ for the ground state and $N_{\mathrm{SVD}}$ for an excited state. They serve as parameters in our method and can be freely adjusted depending on the desired accuracy level and acceptable computational cost. We preface the discussion that follows with a note that, in general, different excited states are expected to have different optimal excitation subspaces for a given dimension $N_{\mathrm{SVD}}$. One may envisage a procedure in which these subspaces are combined into a single one and redundant combinations are dropped. However, in this work we avoid this additional complications and hence the rank-reduced method proposed here is inherently state-specific.

In a nutshell, the strength of the proposed rank-reduced formalism stems from its two major virtues. First, it has been demonstrated numerous times in the literature\cite{parrish19,lesiuk2019,Lesiuk2020,Lesiuk2022,hohenstein19,michalak24} that in order to maintain constant relative error in the energy and other quantities, both for the ground and excited states, the sizes of the triple excitation subspaces have to increase only linearly with the system size. Second, the full-rank amplitudes, i.e. the left-hand sides of Eqs.~(\ref{Tucker1})~and~(\ref{Tucker2}), never have to be explicitly built and stored. As a consequence of these two points, we have to deal only with the three-index objects present on the right-hand sides of Eqs.~(\ref{Tucker1})~and~(\ref{Tucker2}); each index represents a quantity that increases proportionally to the size of the system. Additionally, we point out that the decompositions in Eqs.~(\ref{Tucker1})~and~(\ref{Tucker2}) introduce no \emph{a priori} error. In fact, when the dimensions $N_{\mathrm{svd}}$ and $N_{\mathrm{SVD}}$ are increased to the maximal possible size ($OV$, see Table~\ref{notation}), results of the parent canonical CCSDT/EOM-CCSDT methods are recovered in a continuous fashion. While there is no guarantee that the convergence towards the ``exact'' result is smooth and monotonic, in practice we observe fairly regular decrease of the error, as demonstrated numerically later.

Let us briefly describe the overall workflow of the rank-reduced EOM-CCSDT calculations. The preliminary step of the ground-state/excited-state routine consists of the CCSD/EOM-CCSD calculations which are performed using the conventional algorithm. Using the CCSD/EOM-CCSD result, a guess for the triple excitation amplitudes is implicitly formed using a perturbative approach. Next, the decomposition of the guess amplitudes is carried out, yielding triple-excitation subspaces, i.e. $U_{ia}^x$ or $V_{ia}^X$ in Eqs.~(\ref{Tucker1})~and~(\ref{Tucker2}). Finally, in the last step the CCSDT/EOM-CCSDT equations are solved within the subspace spanned by $U_{ia}^x$/$V_{ia}^X$, giving the compressed amplitudes $t_{xyz}$/$r_{XYZ}$ and an approximate ground-state or excitation energy. In the following sections we describe the latter two steps in detail, focusing on the excited state calculations. For the detailed analysis of the ground-state implementation, see Ref.~\onlinecite{Lesiuk2020} 
\subsection{Guess for the triply-excited amplitudes}

As mentioned in the previous section, a guess for triply-excited amplitudes is required in our formalism to determine the excitation subspaces $U_{ia}^x$/$V_{ia}^X$. For the ground state we follow the procedure used in previous works without any changes\cite{lesiuk2019,Lesiuk2020,Lesiuk2022}. For the excited states we devised two types of guess - a basic and an extended version. We derive them by expanding the triple amplitudes EOM-CCSDT equation in orders of perturbation theory. To this end, we assign orders to individual operators on the basis of the M\o ller-Plesset expansion with the canonical Hartree-Fock orbitals:
\begin{align}
  \tilde{F}_N &\equiv \tilde{F}_N^{(0)}, ~~~~ \tilde{V}_N \equiv \tilde{V}_N^{(1)}, \\
  T_2 &\equiv T_2^{(1)}, ~~~~ T_3 \equiv T_3^{(2)},
\end{align}
where $T_1$ amplitudes are treated as zeroth-order quantities and are absorbed in operators $\tilde{F}_N$ and $\tilde{V}_N$ using the similarity transformation defined previously. Moreover, we aim to accurately describe both singly- and doubly-excited states and hence we treat the $R_1$ and $R_2$ operators on an equal footing by assigning them the zeroth order in the expansion: 
\begin{align}
R_1 \equiv R_1^{(0)}, ~~~~ R_2 &\equiv R_2^{(0)},
\end{align}
whereas the $R_3$ operator is expressed as a series in orders of the perturbation:
\begin{align}
R_3 &= R_3^{(1)} + R_3^{(2)} + \ldots 
\end{align}
This entails a similar expansion for the excited-state triple-excitation amplitudes:
\begin{align}
  R_{ijk}^{abc}= {}^{(1)}R_{ijk}^{abc} + {}^{(2)}R_{ijk}^{abc} + \ldots ,
\end{align}
where the $n$-th order contributions define the corresponding $R_3^{(n)}$ operator.
On the basis of this assignment we consider now the triple amplitudes equation:
\begin{align} \label{triple_amp}
     \langle _{ijk}^{abc}|[\bar{H}_N, R]|0\rangle &= \omega\, \langle _{ijk}^{abc} | R |0\rangle.
\end{align}
To isolate terms of a given order, we write the single-, double- and triple-excitation operators explicitly and evaluate the matrix element on the right hand side of Eq.~\ref{triple_amp}. Additionally, we express the $\bar{H}_N$ operator as an expansion in terms of nested commutators (BCH formula). These operations give us the following form of the previous equation: 
\begin{align} \label{triple_amp_expl}
\nonumber
     &\langle _{ijk}^{abc} | \left[\left( \tilde{F}_N^{(0)} + \tilde{V}_N^{(1)} + [\tilde{F}_N^{(0)}, T_2^{(1)}] + [\tilde{V}_N^{(1)}, T_2^{(1)}] + \ldots \right) , R_1^{(0)} \right] |0\rangle + \\ \nonumber &\langle _{ijk}^{abc} | \left[\left( \tilde{F}_N^{(0)} + \tilde{V}_N^{(1)} + [\tilde{F}_N^{(0)}, T_2^{(1)}] + [\tilde{V}_N^{(1)}, T_2^{(1)}] + \ldots \right) , R_2^{(0)}  \right] |0\rangle = \\ 
       &= \omega ({}^{(1)}R_{ijk}^{abc} + {}^{(2)}R_{ijk}^{abc} + \ldots) - \nonumber\\ 
       &- \langle _{ijk}^{abc} | \left[\left( \tilde{F}_N^{(0)} + \tilde{V}_N^{(1)} + [\tilde{F}_N^{(0)}, T_2^{(1)}] + [\tilde{V}_N^{(1)}, T_2^{(1)}] + \ldots \right) , R_3^{(1)} + R_3^{(2)} + \ldots  \right] |0\rangle.  
\end{align}
Gathering terms according to their order in the perturbation we see that the zeroth-order equation vanishes -- it contains only $\langle _{ijk}^{abc} | [ \tilde{F}_N^{(0)} , R_1^{(0)} ] |0\rangle$ and $\langle _{ijk}^{abc} | [ \tilde{F}_N^{(0)} , R_2^{(0)} ] |0\rangle$ terms which are identically zero. The first non-vanishing contribution is of the first order and reads:
\begin{align}
     \langle _{ijk}^{abc} | [\tilde{V}_N^{(1)}, R_2^{(0)}] |0\rangle =
         \omega \, {}^{(1)}R_{ijk}^{abc} - \langle _{ijk}^{abc} | [\tilde{F}_N^{(0)}, R_3^{(1)} ] |0\rangle.
\end{align}
Evaluation of matrix elements and several rearrangements lead us to the following formula:
\begin{align} \label{basic_guess}
    ^{(1)}R_{ijk}^{abc} = \frac{P_3 \left[(ai\widetilde|lj) R_{lk}^{bc} - (ai\widetilde|bd) R_{jk}^{dc}\right]}{\epsilon_a + \epsilon_b + \epsilon_c - \epsilon_i - \epsilon_j - \epsilon_k - \omega}.
\end{align}
If we consider the second-order terms in Eq.~\ref{triple_amp_expl} we obtain:
\begin{align}
     &\langle _{ijk}^{abc} | \left[[\tilde{V}_N^{(1)}, T_2^{(1)}] , R_1^{(0)} \right] |0\rangle + \langle _{ijk}^{abc} | \left[[\tilde{V}_N^{(1)}, T_2^{(1)}], R_2^{(0)}  \right] |0\rangle  \nonumber \\
     &= \omega \, {}^{(2)}R_{ijk}^{abc} - \langle _{ijk}^{abc} | [F_N^{(0)}, R_3^{(2)}]|0\rangle - \langle _{ijk}^{abc} | [\tilde{V}_N^{(1)} , R_3^{(1)} ]  |0\rangle,
\end{align}
and after some rearrangements we see that the second-order contribution to the triply-excited amplitudes reads:
\begin{align}
     ^{(2)}R_{ijk}^{abc} = - \,\,\frac{ \langle _{ijk}^{abc} | \left[[\tilde{V}_N^{(1)}, T_2^{(1)}] , R_1^{(0)} \right] |0\rangle + \langle _{ijk}^{abc} | \left[[\tilde{V}_N^{(1)}, T_2^{(1)}], R_2^{(0)}  \right] |0\rangle + \langle _{ijk}^{abc} | [\tilde{V}_N^{(1)} , R_3^{(1)} ]  |0\rangle  }{\epsilon_a + \epsilon_b + \epsilon_c - \epsilon_i - \epsilon_j - \epsilon_k - \omega}.
\end{align}
The above formula is rather inconvenient due to the last matrix element, i.e. $\langle _{ijk}^{abc} | [\tilde{V}_N^{(1)} , R_3^{(1)} ]  |0\rangle$, which involves the first-order approximation to the triply-excited amplitudes. Inclusion of this term may be expected to increase the computational costs to such an extent that the calculation would be rendered impractical, at least for larger systems. Therefore, we pragmatically focus only on terms that can be included without a drastic increase of the costs and drop the discussed matrix element. Clearly, this choice is arbitrary and it is unknown at this point what is the relative importance of this term in determination of an appropriate excitation subspace. This issue will be addressed in subsequent sections by analyzing the numerical results obtained with the proposed scheme. Therefore, we consider an approximation to the second-order term defined as:
\begin{align} \label{approx_second}
     ^{(2')}R_{ijk}^{abc} = - \,\,\frac{  \langle _{ijk}^{abc} | \left[[\tilde{V}_N^{(1)}, T_2^{(1)}] , R_1^{(0)} \right] |0\rangle + \langle _{ijk}^{abc} | \left[[\tilde{V}_N^{(1)}, T_2^{(1)}], R_2^{(0)}  \right] |0\rangle  }{\epsilon_a + \epsilon_b + \epsilon_c - \epsilon_i - \epsilon_j - \epsilon_k - \omega},
\end{align}
where the prime in the superscript on the left-hand side of the equation indicates that we use an approximate formula without the $\langle _{ijk}^{abc} | [\tilde{V}_N^{(1)} , R_3^{(1)} ]  |0\rangle$ contribution. Further in the paper, we refer as the \emph{basic guess} to the approximation:
\begin{align}
    ^{(\mathrm{basic})}R_{ijk}^{abc} =~ ^{(1)}R_{ijk}^{abc},
\end{align}
whereas the \emph{extended guess} is defined as the sum of the first- and the approximate second-order terms in the perturbative expansion:
\begin{align}
    ^{(\mathrm{ext})}R_{ijk}^{abc} =~ ^{(1)}R_{ijk}^{abc}  ~+~ ^{(2')}R_{ijk}^{abc}.
\end{align}

\subsubsection{Decomposition of the triply-excited guess amplitudes}

We have now defined the basic and extended guesses for the triply-excited amplitudes for an excited state. However, we never explicitly form these approximate amplitudes in practical calculations. Instead, they are decomposed directly to the form given in Eq.~(\ref{Tucker2}). This is achieved using the higher-order orthogonal iteration procedure (HOOI).\cite{lathauwer2000,kolda09} This method has been previously described in the context of coupled cluster theory in the ground-state\cite{Lesiuk2022} and excited-state\cite{michalak24} calculations, so we limit our discussion to its most essential features and its application in the particular case of basic or extended guess for the amplitudes. 

The HOOI procedure is designed to give the optimal Tucker approximation to a given tensor for a fixed rank $N_{\mathrm{SVD}}$. This is achieved by minimizing the least-squares cost function $f$ which for the particular case of excited-state amplitudes is written as:
\begin{align}
    f = || R_{ijk}^{abc} - r_{XYZ}\,V_{ia}^X\,V_{jb}^Y\,V_{kc}^Z ||^2.
\end{align}
where $||*||$ denotes the Frobenius norm of a tensor. In practice, it is more convenient to recast the above problem as an equivalent maximization of the function $g$:
\begin{align*}
    g = || R_{ijk}^{abc}\,V_{ia}^X\,V_{jb}^Y\,V_{kc}^Z ||^2.
\end{align*}
To find $V_{ia}^X$ that maximizes this expression, the HOOI procedure involves iterative evaluation of the partially compressed tensor:
\begin{align} 
    R_{ia,YZ}^{[n]} = R_{ijk}^{abc} \;^{[n]}V_{jb}^Y\,^{[n]}V_{kc}^Z.
\end{align}
where the superscript $[n]$ denotes the iteration number. In each iteration the $R_{ia,YZ}^{[n]}$ tensor is rearranged as a $OV \times N_{\mathrm{SVD}}^2$ matrix which is then subjected to the singular value decomposition ($\mathrm{SVD}$). The left-singular vectors corresponding to the largest singular values are retained and form the updated tensor $^{[n+1]}V$. The iterations are repeated until the convergence of the triple-excitation subspace. It is important to note that, as a byproduct of the SVD procedure, the column vectors of $V$ are orthonormal in the sense that:
\begin{align} \label{orthonormal}
    V_{ia}^X \, V_{ia}^{X'} = \delta_{XX'}.
\end{align}
We shall use this feature of the subspace later on to simplify some of the working equations. Lastly, note that only the projection ($R_{ia,YZ}$) of the $R_{ijk}^{abc}$ tensor on the subspace $V$ is needed at any given time in the described procedure. This object can be built much more cheaply than the full-rank amplitudes $R_{ijk}^{abc}$ themselves and enables to avoid their storage in memory. Therefore, our next goal is to derive formulas for the projected tensor $R_{ia,YZ}$ in the particular cases of approximate amplitudes from the basic and extended guesses.

In the case of the basic guess this is particularly simple due to large similarity with the guess for the ground-state amplitudes introduced in Ref.~\onlinecite{Lesiuk2022}. In fact, by comparing the formula for the basic guess given in Eq.~\ref{basic_guess} and Eqs~44, 45, 50 in Ref.~\onlinecite{Lesiuk2022}, we see only two minor changes. First, the ground-state amplitudes $T_{ij}^{ab}$ are replaced by their excited-state counterparts $R_{ij}^{ab}$. Second, in the denominator of Eq.~\ref{basic_guess} we have the extra factor of $\omega$ (excitation energy) which is absent in the ground-state formula. As a result, we found that the HOOI procedure introduced for the ground-state amplitudes in Ref.~\onlinecite{Lesiuk2022} is applicable to the excited-state amplitudes by changing only a few lines of code to accommodate the aforementioned differences. Additionally, we point out that the same basic guess for the excited-state amplitudes has been used in the recent work describing the rank-reduced EOM-CC3 calculations.\cite{michalak24} We refer to this paper for further technical details of the HOOI procedure in this case and to Ref.~\onlinecite{Lesiuk2022} for the proof that the computational cost of a single HOOI iteration is proportional to $N^5$.

Application of the HOOI procedure to the amplitudes defined as the extended guess is significantly more involved and below we derive the formula for the projected $R_{ia,YZ}$ tensor in this case. Let us denote the matrix elements appearing in Eq~\ref{approx_second} in the following way:
\begin{align}
    &-\langle _{ijk}^{abc} | \left[[\tilde{V}_N^{(1)}, T_2^{(1)}] , R_1^{(0)} \right] |0\rangle \equiv P_3 \,B_{ijk}^{abc},
    \\
    &-\langle _{ijk}^{abc} | \left[[\tilde{V}_N^{(1)}, T_2^{(1)}], R_2^{(0)}  \right] |0\rangle \equiv P_3 \,C_{ijk}^{abc},
\end{align}
where $P_3$ is the permutation operator defined previously, see Eq.~\ref{p3}, while $B_{ijk}^{abc}$ and $C_{ijk}^{abc}$ are single permutations resulting from the evaluation of the respective matrix elements (with signs reversed for convenience). Similarly, we denote the single-permutation term appearing explicitly in the numerator of Eq.~\ref{basic_guess} as:
\begin{align}
    (ai\widetilde|lj) R_{lk}^{bc} - (ai\widetilde|bd) R_{jk}^{dc} \equiv A_{ijk}^{abc}.
\end{align}
The formula for extended guess takes the form:
\begin{align}
    ^{(\mathrm{ext})}R_{ijk}^{abc} = (\epsilon_{ijk}^{abc} - \omega)^{-1} P_3 \left( A_{ijk}^{abc} + B_{ijk}^{abc} + C_{ijk}^{abc} \right),
\end{align}
where $(\epsilon_{ijk}^{abc} - \omega)^{-1}$ is a denominator from Eqs.~\ref{basic_guess}~and~\ref{approx_second} written in a more compact form. 
Next, we project $^{(\mathrm{ext})}R_{ijk}^{abc}$ in its second and third mode ($jb$ and $kc$) onto the triple-excitation subspace spanned by the tensor $V$:
\begin{align}
   R_{ia,YZ} =~ ^{(\mathrm{ext})}R_{ijk}^{abc} \, V_{jb}^{Y} \, V_{kc}^{Z} = (\epsilon_{ijk}^{abc} - \omega)^{-1}  \, V_{jb}^{Y} \, V_{kc}^{Z} \, P_3  \left( A_{ijk}^{abc} + B_{ijk}^{abc} + C_{ijk}^{abc} \right).
\end{align}
In order to arrive at a tractable form of the above formula we need to eliminate the non-factorizable form of the denominator and find a way to contract the quantities $A_{ijk}^{abc}$, $B_{ijk}^{abc}$ and $C_{ijk}^{abc}$ directly with the tensors spanning the subspace. We resolve the first problem by applying a Laplace transformation\cite{almlof1991} to the denominator:  
\begin{align}
 (\epsilon_{ijk}^{abc} - \omega)^{-1}  = \sum_{g=1}^{N_g} w_g \, e^{-t_g(\epsilon_{ijk}^{abc} - \omega)}
\end{align}
where the sum goes over the grid points of the quadrature with their associated weights $w_g$ and nodes $t_g$. The weights and nodes are determined through the min-max quadrature by Takatsuka et al.\cite{takatsuka2008,braess2005,helmichparis2016} We stress that the size of the quadrature $N_g$ scales as $N^0$, i.e. it does not depend on the system size. The second objective is easily fulfilled using the following identity:
\begin{align}
V_{jb}^{Y} \, V_{kc}^{Z} \, P_3 \left( S_{ijk}^{abc}\right) &= (1 + P_{YZ}) \left[ V_{jb}^{Y} \, V_{kc}^{Z} \left( S_{ijk}^{abc} + S_{jik}^{bac} + S_{kji}^{cba} \right) \right] \\
S_{ijk}^{abc} &= A_{ijk}^{abc} + B_{ijk}^{abc} + C_{ijk}^{abc} 
\end{align}
where, for brevity, we denote the sum of the elements $A_{ijk}^{abc}$, $B_{ijk}^{abc}$ and $C_{ijk}^{abc}$ by $S_{ijk}^{abc}$ and we introduce the permutation operator $P_{YZ}$ which exchanges the indices $Y$ and $Z$. In the last step of the derivation we introduce the scaled projectors:
\begin{align}
   V_{gjb}^{Y} = V_{jb}^Y \, e^{-t_g\epsilon_j^b},
\end{align}
where $\epsilon_j^b = \epsilon_b - \epsilon_j$, which allows us to write the projected tensor $R_{ia, YZ}$ as:
\begin{align} \label{projected_tensor}
   R_{ia,YZ} =~ (1+P_{YZ}) \left[ w_g \, e^{-t_g(\epsilon_{i}^{a} - \omega)} \, V_{gjb}^{Y} \, V_{gkc}^{Z} \left( S_{ijk}^{abc} + S_{jik}^{bac} + S_{kji}^{cba} \right) \right].  
\end{align}
The explicit formula for this tensor alongside its optimal factorization, is derived in Supporting Information. The formal scaling of the resulting equations is $N^6$ which is caused by four terms appearing in the $C_{ijk}^{abc}$ tensor. For brevity, in the following discussion we focus only on these rate-limiting terms. To simplify some steps of the calculations, we introduce the diagonalized ground-state and excited-state double-excitation amplitudes which allow for the optimal factorization:
\begin{align}
 T_{ij}^{ab} = U_{ia}^{'L} \, t_{LM} \, U_{jb}^{'M} \\ 
 R_{ij}^{ab} = V_{ia}^{'W} \, r_{WS} \, V_{jb}^{'S}.
\end{align}
where $t_{LM}$ and $r_{WS}$ are diagonal matrices containing the largest eigenvalues of $T_{ij}^{ab}$ and $R_{ij}^{ab}$, while matrices $U_{ia}^{'L}$ and $V_{ia}^{'W}$ consist of the corresponding eigenvectors. The number of eigenvectors kept in the decomposition (the length of summation over $L,M$ or $W,S$)  of the ground-state and excited-state amplitudes is denoted by $N_{\mathrm{eig}}$ and $N_{\mathrm{EIG}}$, respectively. In the present paper we assume that the following inequalities hold:
\begin{align}
\label{scaling}
 N_{\mathrm{aux}} > N_{\mathrm{eig}} \approx N_{\mathrm{EIG}} > N_{\mathrm{svd}} \approx N_{\mathrm{SVD}} > V \gg O   
\end{align}
which is justified from the point of view of practical calculations.
Additionally, we introduce two intermediates which are needed for the factorization:
\begin{align}
    D_{ia}^{QL} &= \left( B_{ae}^Q \, U_{ie}^{'M} - B_{li}^Q \, U_{la}^{'M} \right) t_{LM}, \\
       F_{ia}^{QW} &= \left( B_{ae}^Q \, V_{ie}^{'S} - B_{li}^Q \, V_{la}^{'S} \right) r_{WS}.   
\end{align}
Notice that upon exchanging the tensors related to the ground-state amplitudes ($U_{ie}^{'M}, U_{la}^{'M}, t_{LM}$) with their excited-state counterparts ($V_{ie}^{'S}, V_{la}^{'S}, r_{WS}$) and \emph{vice versa}, we transform $D_{ia}^{QL}$ and $F_{ia}^{QW}$ intermediates into each other. Due to the analogous symmetry (described below), the four most expensive terms in the evaluation of $R_{ia,YZ}$ come in two pairs which are factorized in the exact same manner. Let us first present the pair of terms with higher scaling:
\begin{align}
    I_1 &= \biggl(\left((D_{ia}^{QL} \, B_{md}^Q) \, U_{mb}^{'L} \right) V_{gjb}^Y \biggl)  \left( V_{jd}^{'W} \left[ (V_{gkc}^Z \, V_{kc}^{'S}) \, r_{WS} \right] \right), \\
    I_2 &= \biggl(\left((F_{ia}^{QW} \, B_{md}^Q) V_{mb}^{'W} \right) V_{gjb}^Y \biggl)  \left( U_{jd}^{'L} \left[ (V_{gkc}^Z \, U_{kc}^{'M}) \, t_{LM} \right] \right).
\end{align}
where the parentheses indicate the optimal order of  contractions. Note that for clarity we do not write the indices on the left-hand sides of the above equations explicitly. The second term of this pair, denoted $I_2$, is obtained from term $I_1$ by exchanging all the tensors corresponding to the ground-state amplitudes for their excited-state counterparts (and \emph{vice versa}), with the exception of the projectors $V_{gjb}^Y$ and $V_{gkc}^Z$ which always refer to the excited-state subspace. Due to the similarity between both terms, we discuss below only the scaling of term $I_1$. This discussion can be easily applied to the other term by taking into account the aforementioned exchange symmetry. The evaluation of term $I_1$ involves four $N^6$ steps, which come from the contractions performed in the left bracket and from the contraction of two outermost brackets with each other. The rest of the terms are much less expensive, with either $N^4$ or $N^3$ scaling.
The precise scaling of the most expensive step is $O^2 V^2 N_{\mathrm{aux}} N_{\mathrm{eig}}$ and is associated with the contraction of $D_{ia}^{QL}$ with $B_{md}^Q$. The remaining $N^6$ steps possess $O^2 V^3 N_{\mathrm{eig}}$, $O^2 V^3 N_{\mathrm{SVD}}$ and $O^2 V^2 N_{\mathrm{SVD}}^2$ scaling. Note that the index $g$ corresponds to the grid points of the quadrature used to evaluate the Laplace transformation numerically. The size of this grid scales as $N^0$ and thus does not rise the formal scaling of the equations. The second pair of terms is related with each other through the same symmetry as described above:
\begin{align}
    I_3 &= \biggl( D_{ia}^{QL} ( B_{md}^Q \, U_{jd}^{'L} ) \biggl) \biggl( V_{gjb}^Y \left( V_{mb}^{'W} \, \left[ (V_{gkc}^Z V_{kc}^{'S}) \, r_{WS} \right] \right) \biggl), \\
    I_4 &= \biggl( F_{ia}^{QW} ( B_{md}^Q \, V_{jd}^{'W} ) \biggl) \biggl( V_{gjb}^Y \left( U_{mb}^{'L} \left[ (V_{gkc}^Z \, U_{kc}^{'M}) \, t_{LM} \right] \right) \biggl).
\end{align}
Focusing on term $I_3$, we see that only two individual contractions scale as $N^6$ which is the highest scaling appearing in this term. Specifically, the rate-limiting step, which involves contraction between $D_{ia}^{QL}$ and the neighboring bracket, scales as $O^3 V N_{\mathrm{aux}} N_{\mathrm{eig}}$. The other step has $O^3 V N_{\mathrm{SVD}}^2$ scaling and comes from the contraction between the outermost brackets. Lastly, we note that the scaling of individual steps of contraction in terms $I_2$ and $I_4$ can be derived from the analysis of terms $I_1$ and $I_3$ by changing $N_{\mathrm{eig}}$ to $N_{\mathrm{EIG}}$, and \emph{vice versa}.

In summary, the projected tensor $R_{ia,YZ}$ is assembled by gathering the $I_1$, $I_2$, $I_3$ and $I_4$ terms, enabling us to rewrite Eq.~\ref{projected_tensor} as:
\begin{align}
   R_{ia,YZ} =~ (1+P_{YZ}) \left[ w_g \, e^{-t_g(\epsilon_{i}^{a} - \omega)} \,  \left( I_1 + I_2 + I_3 + I_4 + \ldots  \right) \right],
\end{align}
where the ellipsis "$\ldots$" denotes the less computationally expensive terms. The extra steps indicated above, i.e. contraction with $w_g \, e^{-t_g(\epsilon_{i}^{a} - \omega)}$ and application of the $(1+P_{YZ})$ operator to the result, involve no additional $N^6$ operations.

To summarize, the overall scaling of the HOOI procedure applied to the extended form of the amplitudes guess is equal to $N^6$, in contrast to the $N^5$ scaling of this procedure when the basic guess is used. The precise scaling of the most expensive terms, expressed through quantities shown in Eq.~\ref{scaling}, is written above, but it is worthwhile to consider the overall scaling with the extended form of the guess in more details. To allow a meaningful analysis we want to approximately relate the values of $N_{\mathrm{aux}}$,  $N_{\mathrm{eig}}$, $N_{\mathrm{EIG}}$, $N_{\mathrm{svd}}$ and $N_{\mathrm{SVD}}$ to the number of virtual orbitals $V$ of a given molecule. The accuracy provided by an auxiliary basis set of size $N_{\mathrm{aux}}=(3-4)\cdot V$ is sufficient for most practical calculations. The sizes of the triple-excitation subspaces for the ground-state $N_{\mathrm{svd}}$ and the chosen excited-state $N_{\mathrm{SVD}}$ which offer a satisfactory accuracy-to-cost ratio are equal to $(1-2)\cdot V$. Next, the numbers of eigenvectors kept in the eigendecomposition of ground-state ($N_{\mathrm{eig}}$) and excited-state ($N_{\mathrm{EIG}}$) double amplitudes are usually somewhat larger than $N_{\mathrm{svd}}$ and $N_{\mathrm{SVD}}$, respectively. In routine calculations $N_{\mathrm{eig}}$ and $N_{\mathrm{EIG}}$ parameters would be typically equal to $(2-4) \cdot V$. To sum up, we use the following approximations for the above parameters:
\begin{align}
    N_{\mathrm{aux}} &\approx (3-4)\cdot V \\
    N_{\mathrm{svd}} &\approx N_{\mathrm{SVD}} \approx (1-2)\cdot V \\
    N_{\mathrm{eig}} &\approx N_{\mathrm{EIG}} \approx (2-4)\cdot V
\end{align}
Now, we reanalyze the scaling of terms $I_1$, $I_2$, $I_3$ and $I_4$. The most expensive contractions for the term $I_1$ have  $O^2V^2N_{\mathrm{aux}}N_{\mathrm{eig}}$,  $O^2V^3N_{\mathrm{eig}}$, $O^2V^3N_{\mathrm{SVD}}$ and $O^2V^2N_{\mathrm{SVD}}^2$ computational costs, which, using the above estimates, sum to the overall $(10-26)\cdot O^2V^4$ scaling. For term $I_3$ the relevant contractions possess $O^3 V N_{\mathrm{aux}} N_{\mathrm{eig}}$ and $O^3 V N_{\mathrm{SVD}}^2$ scaling. In this case, inserting the above estimates gives us the scaling of $(7-20)\cdot O^3V^3$. Taking terms $I_2$ and $I_4$ into account involves simply doubling the above results which gets us to the $(20-52)\cdot O^2V^4 + (14-40)\cdot O^3V^3$ overall scaling. Lastly, although the $N_g$ parameter does not scale with the system size it still contributes to the prefactor. The value of $N_g = 10$ is sufficient for most purposes and gives us the final scaling of the HOOI iterations with extended guess as $(200-520)\cdot O^2V^4 + (140-400)\cdot O^3V^3$. As can be seen from this rough estimate, although the iterations essentially scale as $O^2V^4$ in the leading order, the method is plagued by a substantial prefactor which makes calculations significantly more expensive. From this point of view, using the basic HOOI guess which scales as $N^5$ with the system size is certainly a much more economical endeavor. As demonstrated later there are, however, instances in which the extended guess is significantly more accurate for a given size of the triple-excitation subspace than the basic guess. We will return to this matter when discussing numerical results obtained for both types of guess.

\subsection{Rank-reduced triples residual tensor}

In this section we discuss the factorized triple amplitudes equation as it appears in the rank-reduced formalism. We start from the canonical EOM-CCSDT triples equation (Eq.~\ref{tripleeq}) written in the form:
\begin{align} \label{tripleeq_tensor}
    \Omega_{ijk}^{abc} = \omega R_{ijk}^{abc},
\end{align}
where we have introduced the EOM-CCSDT triples residual tensor $\Omega_{ijk}^{abc}$. In the rank-reduced approach we project each mode of the tensors appearing in Eq.~\ref{tripleeq_tensor} onto the triple-excitation subspace spanned by the tensor $V$, which is obtained from the HOOI iterations. Additionally, we write the $R_{ijk}^{abc}$ amplitudes in the Tucker-3 format. As a result of these operations and after exploiting the orthonormality of the triple-excitation subspace basis (see Eq.~\ref{orthonormal}) we obtain:
\begin{align} \label{tripleeq_tensor2}
    V_{ia}^X \, V_{jb}^Y \, V_{kc}^Z \, \Omega_{ijk}^{abc} = \omega \, r_{XYZ}.
\end{align}
We define the compressed residual tensor $\Omega_{XYZ}=V_{ia}^X \, V_{jb}^Y \, V_{kc}^Z \, \Omega_{ijk}^{abc}$ as the left-hand side of Eq.~\ref{tripleeq_tensor2}, which gives us the final, general formula for the EOM-CCSDT triple amplitudes equation in the rank-reduced approach:
\begin{align}
    \Omega_{XYZ} = \omega \, r_{XYZ}
\end{align}
The benefits of the above procedure are twofold. First, as demonstrated below, we reduce the formal scaling of the method by utilizing the Tucker-3 compression format for the amplitudes and calculating the projected tensor $\Omega_{XYZ}$ instead of the full-rank residual tensor which both allow for efficient factorization of the formulas. Second, the tensors which we store in memory are of size $N_{\mathrm{SVD}}^3$ rather than $O^3V^3$ which significantly reduces the memory requirements of our method. The derivation of the explicit formula for the compressed tensor $\Omega_{XYZ}$ is given in Supporting Information. Here we confine ourselves to the discussion of one of the two rate-limiting terms of the RR-EOM-CCSDT method. Let us start with some general remarks about the derivation which justifies the choice of the presented term. We note that, due to the similarity of obtained terms to the ones which appear in the derivation for the ground-state CCSDT method, part of the equation for the EOM-CCSDT residual tensor is almost completely analogous to the ground-state case~\cite{Lesiuk2020}. It turns out that the most expensive term before the factorization (with $O^3V^5$ scaling, see Supporting Information) is thus factorized exactly like in rank-reduced CCSDT. Because of that, we turn our attention to the term which is ``exclusive'' to the EOM formalism and has the same rate-limiting $OV^5$ scaling as the aforementioned term in the RR-EOM-CCSDT method. This term (subsequently denoted as $E_{ijk}^{abc}$) originates from two matrix elements -- $\langle _{ijk}^{abc} |[[\tilde{V}_N, T_3] , R_1] | 0 \rangle$ and $\langle _{ijk}^{abc} |[[\tilde{V}_N, T_3] , R_2] | 0 \rangle$. It has the following form in the canonical EOM-CCSDT equations:
\begin{align} \label{Ecanonical}
    E_{ijk}^{abc} = - P_3 \left[ \left( (ad\widetilde|le) \, T_{ijk}^{dbe} \right) R_{l}^{c} - \frac{1}{2} \left( (le\widetilde|md)  \, T_{ijk}^{dec} \right) R_{ml}^{ab} \right].
\end{align}
 Evaluation of $E_{ijk}^{abc}$ as indicated by parentheses in Eq.~\ref{Ecanonical} gives $O^4 V^4$ scaling in the rate-limiting step. We now take advantage of the symmetry of triple amplitudes tensor $T_{ijk}^{dec} = T_{ikj}^{dce}$ and subsequently exchange the indices $kc$ and $jb$ in the second term (which is allowed due to the presence of the $P_3$ operator). This gives us:
 \begin{align}
     E_{ijk}^{abc} = - P_3 \left[ \left( (ad\widetilde|le) \, T_{ijk}^{dbe} \right) R_{l}^{c} - \frac{1}{2} \left( (le\widetilde|md) \,  T_{ijk}^{dbe} \right) R_{ml}^{ac} \right].
 \end{align}
 After introducing the following intermediate:
\begin{align}
     \xi_{ad}^{ce} = \left(B_{ad}^Q \, B_{le}^Q\right)R_l^c - \frac{1}{2}\left(B_{le}^Q \, B_{md}^Q \right)R_{ml}^{ac}
\end{align}
we arrive at the compact form for $E_{ijk}^{abc}$:
\begin{align}
    E_{ijk}^{abc} = - P_3 \left( \xi_{ad}^{ce} \, T_{ijk}^{dbe} \right).
\end{align}
One can see that these manipulations actually raised the scaling of this term to the level of $O^3 V^5$, but it turns out that this form provides a good starting point for a factorization in the rank-reduced approach. Namely, in the RR-EOM-CCSDT method the term $E_{ijk}^{abc}$ is projected on the excited-state triple-excitation subspace as indicated by Eq.~\ref{tripleeq_tensor2}. We take advantage of the fact that for an arbitrary tensor $A_{ijk}^{abc}$ we have:
\begin{align}
   V_{ia}^X \, V_{jb}^Y \, V_{kc}^Z \left( P_3 \, A_{ijk}^{abc} \right) = P_{XYZ} \left( V_{ia}^X \, V_{jb}^Y \, V_{kc}^Z  \, A_{ijk}^{abc} \right), \\
   P_{XYZ} = (1 + P_{XY}) (1 + P_{XZ} + P_{YZ}),
\end{align}
where permutation operator $P_{XYZ}$ is completely analogous to the operator $P_3$, but operates on the indices $X$, $Y$ and $Z$.
This allows us to write:
\begin{align}
    E_{XYZ} = - P_{XYZ} \left( V_{ia}^X \, V_{jb}^Y \, V_{kc}^Z \, \xi_{ad}^{ce} \, T_{ijk}^{dbe} \right)
\end{align}
where $E_{XYZ}$ is the projected form of term $E_{ijk}^{abc}$. The final form of this term, which appears in $\Omega_{XYZ}$, is obtained by applying the Tucker decomposition to the ground-state triply-excited amplitudes and manipulating the order of contractions to obtain the optimal factorization. The final, factorized formula reads:
\begin{align}
    E_{XYZ} = - P_{XYZ} \Bigg( V_{ia}^{X} \left[ \left( \big( \xi_{ad}^{ce} \, U_{ke}^{z} \big) \, V_{kc}^Z \right)\bigg( \left( t_{xyz} \, \big( V_{jb}^Y \, U_{jb}^{y} \big) \right) \, U_{id}^{x} \bigg) \right] \Bigg)
\end{align}
Evaluation of $E_{XYZ}$ involves three contractions with the general $N^6$ scaling: the first is $\xi_{ad}^{ce}\,U_{ke}^{z}$, the next is the contraction of the result of the previous step with the $ V_{kc}^Z$ tensor, and the final one is the contraction with the neighboring bracket. This last step is the most expensive with the precise scaling of $OV^2 N_{\mathrm{SVD}}^2 N_{\mathrm{svd}}$. The remaining $N^6$ terms scale as $OV^4 N_{\mathrm{svd}}$ and $OV^3 N_{\mathrm{SVD}} N_{\mathrm{svd}}$. With the estimate $N_{\mathrm{svd}} \approx N_{\mathrm{SVD}} \approx (1-2)\cdot V$, which was used in the discussion of the HOOI procedure, we can roughly approximate the scaling of $E_{XYZ}$ term as $(3-14) \cdot OV^5$. In order to compare the computational costs of RR-EOM-CCSDT and the canonical EOM-CCSDT method we additionally need to take into account also the term which originates from the $O^3 V^5$ expression appearing in the EOM-CCSDT equations. Its factorization is completely analogous to the ground-state case described in Ref.~\onlinecite{Lesiuk2020}, see Supporting Information for explicit formulas. The scaling of this term in the leading order equals $OV^2N_{\mathrm{SVD}}^3 + OV^3 N_{\mathrm{SVD}}^2 + OV^4 N_{\mathrm{SVD}}$. Introducing the above approximation, $N_{\mathrm{SVD}} \approx (1-2) \cdot V$, the estimated scaling in terms of the number of occupied and virtual orbitals is $(3-14) \cdot OV^5$, which is the same as for the ``EOM exclusive'' term described above. As we can see, the RR-EOM-CCSDT method presented here is approximately twice as expensive as the ground-state rank-reduced CCSDT it is based on. Summing up the costs of both terms, we get the overall leading-order scaling of the RR-EOM-CCSDT method equal $(6-28) \cdot OV^5$. It is worthwhile to compare it to the canonical EOM-CCSDT calculations. The rate-determining step for the canonical method has $O^3 V^5$ scaling. Thus, the ratio of the cost of RR-EOM-CCSDT method to its canonical counterpart is $\frac{(6-28) \cdot OV^5}{O^3V^5}=\frac{6-28}{O^2}$. This shows that even in the worst-case scenario the computational cost of the canonical EOM-CCSDT method in the function of the system size rises above the RR-EOM-CCSDT cost relatively early on -- for systems having roughly 6 or more occupied orbitals. Of course, this estimate does not take into account the EOM-CCSD calculations or the HOOI procedure which need to be performed before the RR-EOM-CCSDT iterations in the rank-reduced approach. The HOOI iterations with the extended form of guess have significant impact on the overall computational costs. Nevertheless, the avoidance of the $N^8$ scaling of the canonical method is crucial for the larger systems, for which the cost of performing the EOM-CCSDT iterations should far exceed the cost associated with RR-EOM-CCSDT and all the preliminary calculations. 

\section{Results and discussion}

\subsection{Computational details}
We tested the performance of the RR-EOM-CCSDT method in a series of calculations for various systems using both basic and extended types of guess. These calculations consist of three distinct types: (i) benchmarks for individual excited states with varying excitation character for a selected group of molecules, (ii) calculations of excited-state potential energy curves (PECs) for magnesium dimer and (iii) investigation of intermolecular charge-transfer excitation in the $\mathrm{NH}_3$-$\mathrm{F}_2$ complex.

Let us first provide a couple of general technical remarks which apply to all of the aforementioned applications before we focus on specific systems listed above. As previously mentioned, the first step of our excited-state routine consists of EOM-CCSD iterations. This preliminary calculation utilizes the block version of Davidson's algorithm\cite{davidson1975,liu1978}, yielding several lowest excited states. On the other hand, in the RR-EOM-CCSDT iterations a single-root version of this algorithm is used and a preselected converged EOM-CCSD vector is followed.\cite{butscher1976} In general, the thresholds for convergence of Davidson's algorithm, both in block and single-root steps, were set to $10^{-5}$ for the energy and $10^{-4}$ for the value of the maximum component of the residual vector. There were, however, two exceptions for which we used lower thresholds. First, in the calculations for nitroxyl and water molecules as a function of $N_{\mathrm{SVD}}$, the thresholds in the single-root step of the routine (RR-EOM-CCSDT iterations) were by an order of magnitude smaller than given above in order to ensure that the incompleteness of the triple-excitation subspace size is the sole source of error. The second exception is the calculation of the asymptotic energies of magnesium dimer for a large interatomic distance, where the thresholds for the energy and residual vector were set to $10^{-7}$ and $10^{-6}$, respectively, for both the block and single-root steps of the routine. In this case, smaller thresholds were adopted to avoid propagation of errors in the evaluation of the interaction energies for each state.

In each calculation, the triple-excitation subspace for an excited state was obtained through the HOOI procedure with either the basic or extended guess, while the procedure from Ref.~\onlinecite{Lesiuk2022} was adopted for the ground state. The dimensions of the excitation subspaces for the ground state and an excited state, denoted as $N_{\mathrm{svd}}$ and $N_{\mathrm{SVD}}$ {(see Table~\ref{notation})}, are the critical parameters in the proposed formalism. Following Ref.~\onlinecite{michalak24} we set $N_{\mathrm{svd}}=N_{\mathrm{SVD}}$ in all calculations reported here and hence the symbol $N_{\mathrm{SVD}}$ refers to both subspace sizes for simplicity. Next, we relate the parameter $N_{\mathrm{SVD}}$ to the number of active molecular orbitals of a given system, $N_{\mathrm{MO}}$ (excluding the frozen-core orbitals). In the case when $N_{\mathrm{SVD}}$ would not be an integer (for example, subspace size of $1/2\cdot N_{\mathrm{MO}}$ and an odd value of $N_{\mathrm{MO}}$) the subspace sizes were always rounded up to the next whole number.

To speed up the HOOI procedure we employ the diagonalized representation of the $T_{ij}^{ab}/R_{ij}^{ab}$ amplitudes. This representation is subsequently truncated by dropping all eigenpairs with eigenvalues smaller in absolute value than a threshold $\sigma_{\mathrm{thr}}$. The particular values of $\sigma_{\mathrm{thr}}$ used in each part of the calculations are given in the corresponding discussion. Lastly, in all calculations we use the frozen-core approximation -- the 1s orbitals of all atoms were uncorrelated.

\subsection{Benchmark calculations}
In this part of the calculations we investigated isolated excited states of several molecules, namely: acrolein, butadiene, benzene, nitrosomethane, nitroxyl, glyoxal, water, acetylene, tetrazine and cyclobutadiene. The optimized geometries of all molecules were taken from the literature. In the case of $\mathrm{H}_2\mathrm{O}$ and $\mathrm{C}_2\mathrm{H}_2$ we use geometries reported in Ref.~\onlinecite{michalak24}. For tetrazine and cyclobutadiene we utilize geometries reported by Kossoski et al.\cite{kossoski2024}, while the geometries of the remaining systems were taken from the article by Loos et al\cite{loos2022}.  The threshold for the eigendecomposition of $T_{ij}^{ab}$ and $R_{ij}^{ab}$ was equal $\sigma_{\mathrm{thr}} = 5 \cdot 10^{-5}$. The calculations were carried out using Dunning-type basis sets, aug-cc-pVDZ and aug-cc-pVTZ\cite{dunning1989a,kendall1992a}, with the purpose of testing the RR-EOM-CCSDT method against the canonical EOM-CCSDT theory. In the calculations we used both basic and extended forms of guess with several triple-excitation subspace sizes. The results are gathered in Tables~\ref{DZ_basic_guess}, \ref{DZ_extended_guess}, \ref{TZ_basic_guess} and \ref{TZ_extended_guess}, each corresponding to a different basis and guess type combination. 

The investigated excited states differ in character, as indicated by the $\%R_1$ parameter which is calculated as a square norm of the $R_i^a$ amplitudes.\footnote{The total RR-EOM-CCSDT eigenvectors are normalized to the unity -- we emphasize that $R_{ij}^{ab}$ and $r_{XYZ}$ are not limited to unique excitations in calculation of the norm.} Note that values of the $\%R_1$ parameter for a given molecule and excited state vary only slightly with the basis set and type of guess (the largest difference is 1.3 percentage points for acrolein in Tables 2 and 5). Among the studied states, the $\%R_1$ parameter is the largest for the first excited state of acetylene ($\approx 90$-$91$\%) and water ($\approx 86$\%), indicating that both states can be described as dominated by single excitations. On the other end of the spectrum, we have genuine higher-excited states of nitrosomethane, nitroxyl, glyoxal, tetrazine and cyclobutadiene. For these states the $\%R_1$ parameter ranges from 0.0 to 0.2\% and hence they have predominantly double-excitation character as the $R_{ij}^{ab}$ and $r_{XYZ}$ contributions to the norm are in the ratio of roughly 70:30, and up to around 77:23, in favor of double excitations. The remaining excited states of acrolein, benzene and butadiene possess an intermediate character between the single- and higher-excitations. The exact EOM-CCSDT excitation energies are gathered in the third column of Tables~\ref{DZ_basic_guess}, \ref{DZ_extended_guess}, \ref{TZ_basic_guess} and \ref{TZ_extended_guess}. They were taken from articles by Loos et al.\cite{loos2022} and Kossoski et al.\cite{kossoski2024}, except for water and acetylene excitation energies, which we calculated using the CFOUR program\cite{cfour,cfour2}. In the CFOUR calculations the thresholds for convergence of SCF, CC (using ECC program) and excited-state CC calculations were set to $10^{-10}$, $10^{-10}$ and $10^{-8}$, respectively. In the next four columns we report errors in excitation energies, i.e. differences between the RR-EOM-CCSDT and the exact EOM-CCSDT results, $\Delta \omega = \omega_{\mathrm{RR-EOM-CCSDT}} - \omega_{\mathrm{EOM-CCSDT}}$. The errors are presented for four triple-excitation subspace sizes: 0.5, 1.0, 1.5 and 2.0 $N_{\mathrm{MO}}$. In the last column we present the absolute values of the ratio $\Delta \omega /\omega_{R_3}$ (calculated for $N_{\mathrm{SVD}} = 2N_{\mathrm{MO}}$), where $\omega_{R_3}$ is the difference between the exact EOM-CCSDT excitation energy and the exact EOM-CCSD energy for a given state. The value of |$\Delta \omega /\omega_{R_3}$| tells us how the accuracy of our method compares with the accuracy of EOM-CCSD or, equivalently, how well the method captures the pure contribution of triple excitations to the excitation energy. Notice that the value of \%|$\Delta \omega /\omega_{R_3}$| = 100 \% would mean that our method is effectively indistinguishable in terms of accuracy from the standard EOM-CCSD procedure for a given system, as it would give the same absolute error with respect to the canonical EOM-CCSDT.

We preface the discussion of the numerical results with a couple of remarks on accuracy of EOM-CCSDT. The mean absolute error (MAE) of the EOM-CCSDT method with respect to FCI for singly-excited states equals about 0.03 eV.\cite{loos2018} On the other hand, the EOM-CCSDT method gives considerably larger errors for states with significant contribution of double (and higher) excitations. In cases where this contribution is moderate, see Ref.~\onlinecite{kossoski2024} for precise definitions, the MAE of EOM-CCSDT with respect to the theoretical best estimate grows to 0.09 eV. For states dominated by double excitations, the corresponding error is 0.42 eV~\cite{kossoski2024}. In what follows, we discuss the RR-EOM-CCSDT numerical results in the context of the above MAEs of EOM-CCSDT. The acceptable errors resulting from the rank reduction should be at least several times smaller than the inherent error of the EOM-CCSDT method for a given problem. This means that somewhat larger errors are acceptable for intermediate states and especially for states of genuine double-excitation character. Nonetheless, to make the proposed method useful in practice, we still expect the rank-reduced method to maintain accuracy of a few hundredth parts of eV (with respect to the canonical counterpart) even for the most problematic cases.
 
We now turn our attention to Tables~\ref{DZ_basic_guess}~and~\ref{DZ_extended_guess} which contain aug-cc-pVDZ results calculated with basic and extended guesses, respectively. In the smallest subspace size ($N_{\mathrm{SVD}} = 0.5~N_{\mathrm{MO}}$) the mean absolute errors in excitation energies ($|\Delta \omega|$) are large and amount to 0.070 eV for the basic guess and 0.103 eV for the extended guess.
Some smaller errors are sporadically encountered, e.g. for nitrosomethane calculation with basic guess, but these are likely accidental as indicated by a significant increase of the error in a larger subspace. The errors for states dominated by higher than single excitations are particularly large, especially for the extended type of guess. Exceptionally large errors are also encountered for the intermediate state of acrolein for both basic and extended guess types. Thus, the calculations in the smallest subspace size ($N_{\mathrm{SVD}} = 0.5~N_{\mathrm{MO}}$) fail to reproduce the EOM-CCSDT results to the desired accuracy level.

Moving to the results obtained for larger subspace sizes we see a significant decrease in errors. Considering the $N_{\mathrm{SVD}} = 1~N_{\mathrm{MO}}$ case next, the mean absolute errors in excitation energies drop to 0.033 eV, both for the basic guess and extended guess. This level of error is still large; moreover, equally worryingly we find a considerable spread of the errors. For example, the error for the nitroxyl molecule with the basic guess is $-0.119$ eV, while it is nearly zero for cyclobutadiene. Similar conclusions hold for the extended guess -- for example, the error for glyoxal reaches 0.086 eV, while the error for nitrosomethane equals $-0.005$ eV. Therefore, the subspace size $N_{\mathrm{SVD}} = 1 N_{\mathrm{MO}}$ is not sufficient to obtain reliable results with the RR-EOM-CCSDT method.

Next, we turn our attention to the results obtained for the subspace size of $N_{\mathrm{SVD}} = 1.5~N_{\mathrm{MO}}$. The average absolute errors further decrease and amount to 0.018 and 0.014 eV for the basic and extended guess, respectively. The average errors are thus approximately 1.5-2.0 times smaller than the inherent error of EOM-CCSDT for singly-excited states. Equally importantly, we see that the maximum errors obtained with $N_{\mathrm{SVD}} = 1.5~N_{\mathrm{MO}}$ are considerably reduced in comparison with $N_{\mathrm{SVD}} = 1~N_{\mathrm{MO}}$. The largest absolute error is encountered for nitroxyl (0.060 eV) for the basis guess and acrolein (0.043 eV) for the extended guess. In light of this observations we can conclude that the $1.5~N_{\mathrm{MO}}$ results calculated with either guess are considerably improved, although some of the individual errors are still larger than desired.

This problem is solved by moving to the $N_{\mathrm{SVD}} = 2.0~N_{\mathrm{MO}}$ subspace size in which the mean absolute errors are 0.009 eV for the basic guess and 0.008 eV for the extended guess. Simultaneously, the maximum errors drop to the level of 0.026 eV for the basic guess and 0.024 eV for the extended guess. In particular, if we restrict the discussion to states dominated by the single excitations, the largest error is obtained for the acetylene molecule which is roughly two and five times smaller than the inherent EOM-CCSDT error for this type of states with the basic and extended guess, respectively. For intermediate states and states dominated by double excitations, the mean absolute errors are uniformly at least several times smaller than the aforementioned inherent EOM-CCSDT errors with respect to FCI. Moreover, these errors do not exceed a few hundredth parts of eV. Therefore, $N_{\mathrm{SVD}} = 2.0~N_{\mathrm{MO}}$ is the smallest subspace size which satisfies our accuracy goals. We expect the errors to decrease when the parameter $N_{\mathrm{SVD}}$ is enlarged further, but mean errors significantly below 0.01 eV are probably necessary only in very accurate calculations for small systems, rather than in routine applications. We hence recommend $N_{\mathrm{SVD}} = 2.0~N_{\mathrm{MO}}$ as default value for typical calculations.

It is also interesting to measure the errors obtained for $N_{\mathrm{SVD}} = 2.0~N_{\mathrm{MO}}$ in relation to the contribution of triple excitations to the excitation energy. The last column in Tables 2 and 3 shows values of  \%|$\Delta \omega /\omega_{R_3}$|
calculated for 2$N_{\mathrm{MO}}$ subspace size with basic and extended guess, respectively. We see that the pure triple-amplitudes contribution to the excitation energy is reproduced accurately for the vast majority of systems and both types of guess. In fact, the error resulting from the tensor decomposition constitutes more than 5\% of the $\omega_{R_3}$ value only for acetylene. In cases where the triply-excited contribution to the excitation energy is large, especially for states dominated by double excitations, the parameter \%|$\Delta \omega /\omega_{R_3}$| never exceeds 1\%. This shows that RR-EOM-CCSDT captures the triple-excitation effects faithfully even for states where the EOM-CCSD method is wrong by several eV, despite the fact that the guess used for determination of the triple-excitation subspace is based entirely on the EOM-CCSD amplitudes.

\begin{table}[H]
\centering
 \begin{tabular}{M{3cm}|S[table-format=2.1]|M{2.2cm}|S[table-format=-1.4]|S[table-format=-1.4]|S[table-format=-1.4]|S[table-format=-1.4]|S[table-format=2.2]}
 \hline
    \multicolumn{7}{M{14cm}}{aug-cc-pVDZ (basic guess)} \\
    \hline
   \multirow{2}{*}{molecule}    & {\multirow{2}{*}{$\%R_1$}} & \multirow{2}{*}{ex. CCSDT$^a$} &\multicolumn{3}{M{4cm}}{\;\;\;\;$\Delta \omega$} & &{\multirow{2}{*}{$\%|\Delta \omega / \omega_{R_3}|$}}\\
           \cline{4-7}
         &  &
           & {0.5 $N_{\mathrm{MO}}$}& 
           {1.0 $N_{\mathrm{MO}}$} & 
           {1.5 $N_{\mathrm{MO}}$} & 
           {2.0 $N_{\mathrm{MO}}$} &  \\
         \hline
         acrolein  &55.2
  &6.717 &0.159 &0.050 & 0.022 & 0.014 & 2.9
\\
butadiene  &43.7 &6.589 &0.066 &0.017 & 0.008 & 0.002 & 0.3
\\
benzene  &70.9 & 5.083 & 0.034 & 0.008 & -0.003& -0.005 & 4.0

\\
nitrosomethane  & 0.2 &5.258 &-0.004 
&-0.057& -0.024& -0.005 & 0.1
 \\
nitroxyl   & 0.1 & 4.756 & -0.103
& -0.119& -0.060& -0.026 & 0.7

\\
glyoxal  & 0.0 & 6.222 & 0.117 
& 0.014 & -0.005 & 0.007 & 0.1

\\
water  &85.6 &7.482 &0.037 
&-0.003 & 0.005 & -0.002 & 4.9

\\
acetylene  &90.9 &7.323 &0.006 &-0.045 
& -0.026& -0.014 & 24.2

\\
tetrazine  &0.1 &5.858 & 0.124& -0.018 & -0.023 & -0.009 & 0.2
\\
cyclobutadiene &0.2 &4.327 &0.045&0.000& -0.002& 0.003 & 0.1

\\ \hline
MAE & {--} & {--} & 0.070 & 0.033 & 0.018 & 0.009  & {--}
    \end{tabular}
    \caption{Comparison between the RR-EOM-CCSDT method with the basic guess and the canonical EOM-CCSDT in the aug-cc-pVDZ basis set. Exact excitation energies and the calculated errors are given in eV.}
    \label{DZ_basic_guess}
    {\noindent\footnotesize\linespread{1.0}\selectfont \begin{flushleft}  $^a$the exact EOM-CCSDT excitation energies of water and acetylene were calculated with CFOUR program. Excitation energies of the remaining molecules were taken from Ref.~\onlinecite{kossoski2024} (tetrazine and cyclobutadiene) and Ref.~\onlinecite{loos2022} (other systems).\end{flushleft} }
    \end{table} 
    
\begin{table}[H]
\centering
 \begin{tabular}{M{3cm}|S[table-format=2.1]|M{2.2cm}|S[table-format=-1.4]|S[table-format=-1.4]|S[table-format=-1.4]|S[table-format=-1.4]|S[table-format=2.2]}
 \hline
   \multicolumn{7}{M{14cm}}{aug-cc-pVDZ (extended guess)} \\
    \hline
   \multirow{2}{*}{molecule}    & {\multirow{2}{*}{$\%R_1$}} & \multirow{2}{*}{ex. CCSDT$^a$} &\multicolumn{3}{M{4cm}}{\;\;\;\;$\Delta \omega$} & &{\multirow{2}{*}{$\%|\Delta \omega / \omega_{R_3}|$}}\\
           \cline{4-7}
         &  &
           & {0.5 $N_{\mathrm{MO}}$}& 
           {1.0 $N_{\mathrm{MO}}$} & 
           {1.5 $N_{\mathrm{MO}}$} & 
           {2.0 $N_{\mathrm{MO}}$} &  \\
         \hline
         acrolein  &55.7
  &6.717 &0.205 &0.081 & 0.043 & 0.020 & 4.1

\\
butadiene  &44.0
  &6.589 &0.078 &0.026
 & 0.008 & 0.004 & 0.8

\\
benzene  & 71.0
   & 5.083 & 0.045 & 0.019
 & 0.006 & 0.000 & 0.2

\\
nitrosomethane  & 0.2
  &5.258 &0.047 &-0.005
& 0.005 & -0.001 & 0.0

 \\
nitroxyl   & 0.1
  & 4.756 & 0.036 & -0.010
& -0.008 & -0.003 & 0.1

\\
glyoxal  & 0.0
  & 6.222 & 0.217
& 0.086 & 0.028 & 0.024 & 0.5

\\
water  &85.6
  &7.482 & 0.048
&0.009 & 0.005 & -0.001 & 2.7

\\
acetylene  &90.8
  &7.323 & 0.003
&-0.017 & -0.009 & -0.006 & 10.8

\\
tetrazine  &0.1
  &5.858 & 0.221
&0.036 & 0.008 & 0.004 & 0.1

\\
cyclobutadiene &0.2
  &4.327 & 0.126
&0.038 & 0.023 & 0.016 & 0.5

\\ \hline
MAE & {--} & {--} & 0.103 & 0.033  & 0.014  & 0.008 & {--}
    \end{tabular}
    \caption{Comparison between the RR-EOM-CCSDT method with the extended guess and the canonical EOM-CCSDT in the aug-cc-pVDZ basis set. Exact excitation energies and the calculated errors are given in eV.}
    \label{DZ_extended_guess}
    {\noindent\footnotesize\linespread{1.0}\selectfont \begin{flushleft}  $^a$the exact EOM-CCSDT excitation energies of water and acetylene were calculated with CFOUR program. Excitation energies of the remaining molecules were taken from Ref.~\onlinecite{kossoski2024} (tetrazine and cyclobutadiene) and Ref~\onlinecite{loos2022} (other systems). \end{flushleft} }
    \end{table} 

Let us now move to the results obtained in aug-cc-pVTZ basis set which are presented in Tables~\ref{TZ_basic_guess}~and~\ref{TZ_extended_guess}. The conclusions from calculations in both basis sets are very similar and the results follow the same general trend. Therefore, in this paragraph we focus on the main differences and provide an overview. First, note that the mean absolute errors calculated for a given subspace size and guess combination are always smaller than their aug-cc-pVDZ counterparts, although these differences are rather small, ranging between 0.001 and 0.023 eV. Similar conclusions hold for the errors expressed through the quantity \%|$\Delta \omega /\omega_{R_3}$|. The maximum absolute errors also decrease when going from aug-cc-pVDZ to aug-cc-pVTZ basis, with the only notable exception being calculations with extended type of guess for $N_{\mathrm{SVD}} = 2.0~N_{\mathrm{MO}}$, where the maximum error obtained within aug-cc-pVTZ basis set is somewhat larger. Therefore, we stand by our recommendation to use $N_{\mathrm{SVD}} = 2.0~N_{\mathrm{MO}}$ as the default size of the triple excitation subspace for both basis sets. Although extension of this subspace to even larger sizes is possible and should further reduce the error related to the tensor decomposition, it may prove to be redundant as the basis set incompleteness error and various other approximations can spoil the overall accuracy of the result.

\begin{table}[H]
\centering
 \begin{tabular}{M{3cm}|S[table-format=2.1]|M{2.2cm}|S[table-format=-1.4]|S[table-format=-1.4]|S[table-format=-1.4]|S[table-format=-1.4]|S[table-format=2.2]}
 \hline
    \multicolumn{7}{M{14cm}}{aug-cc-pVTZ (basic guess)} \\
    \hline
   \multirow{2}{*}{molecule}    & {\multirow{2}{*}{$\%R_1$}} & \multirow{2}{*}{ex. CCSDT$^a$} &\multicolumn{3}{M{4cm}}{\;\;\;\;$\Delta \omega$} & &{\multirow{2}{*}{$\%|\Delta \omega / \omega_{R_3}|$}}\\
           \cline{4-7}
         &  &
           & {0.5 $N_{\mathrm{MO}}$}& 
           {1.0 $N_{\mathrm{MO}}$} & 
           {1.5 $N_{\mathrm{MO}}$} & 
           {2.0 $N_{\mathrm{MO}}$} &  \\
         \hline
         acrolein  &56.2
  &6.730 &0.109 &0.033
 & 0.014 & 0.008 & 1.5
\\
butadiene  &44.7
  &6.598 &0.053 &0.013
 & 0.003 & -0.001 & 0.2

\\
benzene  & 71.2   & 5.062 & 0.025 & 0.001
 & -0.008 & -0.004 & 3.0

\\
nitrosomethane  & 0.2
  &5.293 &-0.059 &-0.033
& -0.007 & 0.001 & 0.0

 \\
nitroxyl   &0.1
  &4.785 & -0.078 & -0.075
& -0.043 & -0.016 & 0.4

\\
glyoxal  & 0.0
  & 6.353 & 0.099 & 0.007
 & 0.015 & 0.014 & 0.2

\\
water  &85.5
  &7.575 & 0.015 &-0.001
 & -0.002 & 0.000 & 6.4

\\
acetylene  &90.4
 &7.209 & -0.038 &-0.018
 & -0.008 & -0.005 & 8.8

\\
tetrazine  &0.1
  &5.958 & 0.070 &-0.010
 & -0.007 & -0.002 & 0.0

\\
cyclobutadiene &0.2
  &4.429 & 0.038 &0.003
 & -0.005 & 0.001 & 0.0

\\ \hline
MAE & {--} & {--} & 0.059  & 0.019 & 0.011 & 0.005 & {--}
    \end{tabular}
    \caption{Comparison between the RR-EOM-CCSDT method with the basic guess and the canonical EOM-CCSDT in the aug-cc-pVTZ basis set. Exact excitation energies and the calculated errors are given in eV.}
    \label{TZ_basic_guess}
    {\noindent\footnotesize\linespread{1.0}\selectfont \begin{flushleft} $^a$the exact EOM-CCSDT excitation energies of water and acetylene were calculated with CFOUR program. Excitation energies of the remaining molecules were taken from Ref.~\onlinecite{kossoski2024} (tetrazine and cyclobutadiene) and Ref~\onlinecite{loos2022} (other systems). \end{flushleft}}
    \end{table}  

\begin{table}[H]
\centering
 \begin{tabular}{M{3cm}|S[table-format=2.1]|M{2.2cm}|S[table-format=-1.4]|S[table-format=-1.4]|S[table-format=-1.4]|S[table-format=-1.4]|S[table-format=2.2]}
 \hline
    \multicolumn{7}{M{14cm}}{aug-cc-pVTZ (extended guess)} \\
    \hline
   \multirow{2}{*}{molecule}    & {\multirow{2}{*}{$\%R_1$}} & \multirow{2}{*}{ex. CCSDT$^a$} &\multicolumn{3}{M{4cm}}{\;\;\;\;$\Delta \omega$} & &{\multirow{2}{*}{$\%|\Delta \omega / \omega_{R_3}|$}}\\
           \cline{4-7}
         &  &
           & {0.5 $N_{\mathrm{MO}}$}& 
           {1.0 $N_{\mathrm{MO}}$} & 
           {1.5 $N_{\mathrm{MO}}$} & 
           {2.0 $N_{\mathrm{MO}}$} &  \\
         \hline
         acrolein  &56.5
  &6.730 &0.144
 &0.050 & 0.021 & 0.013 & 2.3

\\
butadiene  &44.7
  &6.598 &0.064
 &0.017 & 0.005 & 0.002 & 0.4

\\
benzene  & 71.2   & 5.062 & 0.028
 & 0.008 & -0.003 & -0.001 & 0.6

\\
nitrosomethane  & 0.2
  &5.293 &0.019
 &0.004 & 0.009 & 0.004 & 0.1

 \\
nitroxyl   &0.1
  &4.785 & 0.081
 & -0.011 & 0.004 & 0.006 & 0.1

\\
glyoxal  &0.0
  &6.353 & 0.204
 &0.053
& 0.040 & 0.026 & 0.4

\\
water  &85.6
 &7.575 & 0.033
&-0.001 & -0.001 & 0.000 & 1.2

\\
acetylene  &90.4
 &7.209 & 0.014
 &-0.002 & -0.004 & -0.002 & 3.6

\\
tetrazine  &0.1
  &5.958 & 0.134 &0.028
& 0.012 & 0.009 & 0.1

\\
cyclobutadiene &0.2
  &4.429 & 0.079
 &0.030 & 0.008 & 0.006 & 0.2

\\ \hline
MAE & {--} & {--} & 0.080 & 0.020 & 0.011 & 0.007 & {--}
    \end{tabular}
    \caption{Comparison between the RR-EOM-CCSDT method with the extended guess and the canonical EOM-CCSDT in the aug-cc-pVTZ basis set. Exact excitation energies and the calculated errors are given in eV.}
    \label{TZ_extended_guess}
    {\noindent\footnotesize\linespread{1.0}\selectfont \begin{flushleft}  $^a$the exact EOM-CCSDT excitation energies of water and acetylene were calculated with CFOUR program. Excitation energies of the remaining molecules were taken from Ref.~\onlinecite{kossoski2024} (tetrazine and cyclobutadiene) and Ref~\onlinecite{loos2022} (other systems). \end{flushleft} }
    \end{table}  

\begin{figure}[H]
    \centering
    \includegraphics[scale=1.0]{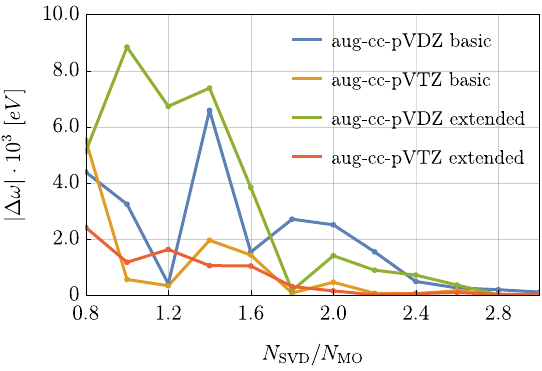}
    \caption{\footnotesize Absolute error in the first excitation energy of water (RR-EOM-CCSDT vs. canonical EOM-CCSDT) plotted against the triple-excitation subspace size expressed in units of the number of molecular orbitals $N_{\mathrm{MO}}$ of the system. The results are presented for four different combinations of basis set and type of guess.}
    \label{h2o_errors}
\end{figure}

\begin{figure}[H]
    \centering
    \includegraphics[scale=1.0]{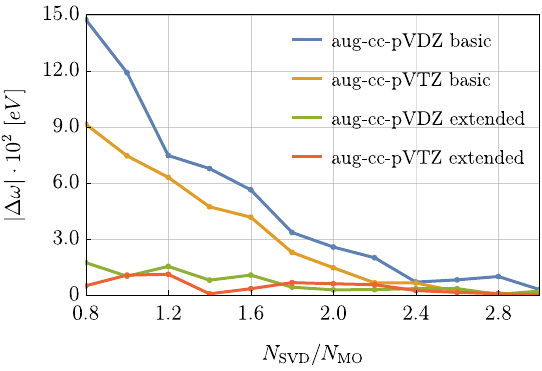}
    \caption{\footnotesize  
    Absolute error in the excitation energy of the lowest-lying excited state of nitroxyl molecule with dominant contribution of double excitations (RR-EOM-CCSDT vs. canonical EOM-CCSDT) plotted against the triple-excitation subspace size expressed in units of the number of molecular orbitals $N_{\mathrm{MO}}$ of the system. The results are presented for four different combinations of basis set and type of guess.}
    \label{nitroxyl_errors}
\end{figure}

Lastly, we address the problem of choosing the type of guess. In order to give a recommendation in this matter we must take a more practical viewpoint and compare both the accuracy and computational costs of the basic- and extended-guess approaches. Regarding the overall accuracy, both types of guess give similar results, as described in the preceding discussion. However, their accuracy for a given system can be widely different. In Figs.~\ref{h2o_errors}~and~\ref{nitroxyl_errors} we present the absolute errors in excitation energies for the first excited state of water and the first genuine higher-excitation of nitroxyl, respectively. In regards to water calculations, we compare the RR-EOM-CCSDT results against our canonical EOM-CCSDT calculations in the CFOUR program, while the exact EOM-CCSDT excitation energies of nitroxyl were taken from article by Loos et al\cite{loos2019b}. The errors are plotted against the subspace size in the units of $N_{\mathrm{MO}}$, giving us a more detailed look on the convergence to the exact result for two states of different character and for four distinct combinations of basis set and guess type. We note that in these calculations the $T_{ij}^{ab}$/$R_{ij}^{ab}$ amplitudes utilized in the HOOI procedure were not approximated in order to ensure that the observed errors stem only from the incompleteness of the triple-excitation subspace size. 

In the case of water molecule, see Fig.~\ref{h2o_errors}, all errors lie below 0.01 eV. The relative accuracy of basic and extended guess calculations changes significantly with the subspace size. In the investigated range of subspaces we see relatively large fluctuations of error in comparison to its absolute value. It is especially pronounced for aug-cc-pVDZ basis set, where change of the subspace size by a relatively small factor of 0.2 $N_{\mathrm{MO}}$ may change the observed error significantly. Of course, these fluctuations are more pronounced for smaller subspace sizes, up to 2.0 $N_{\mathrm{MO}}$. In larger subspaces we see clear convergence to the exact result. In the recommended 2.0 $N_{\mathrm{MO}}$ subspace the extended guess is more accurate than the basic guess in a given basis, although the absolute differences in the error are minor. Thus, in the case of water molecule the accuracy of both types of guess is comparable.  

We now turn our attention to Fig.~\ref{nitroxyl_errors}. In this case the fluctuation of error with the growing subspace size is small and we observe that the error decreases rather steadily for calculations with basic guess in either basis set or maintains an approximately constant value when the extended version of guess is used. Most importantly, the extended guess is much more accurate in this case for the smaller subspace sizes and remains more accurate in the recommended 2.0 $N_{\mathrm{MO}}$ subspace. Further extension of the subspace size reduces these differences and for 3.0 $N_{\mathrm{MO}}$ subspace both types of guess are practically equivalent in both basis sets.

The above examples show one system for which the basic- and extended-guess accuracy is comparable and one for which the performance of the extended guess is much better. We see that in these cases the extended guess provides high-quality results for states of single- and higher-excitation character despite the fact that one of the second-order terms which should be included in its definition has been discarded. However, looking beyond the results obtained for water and nitroxyl molecules reveals that the extended guess in many cases exhibits larger errors than the basic guess. For example, this is observed for glyoxal, acrolein and cyclobutadiene in both basis sets. Additionally, the extended guess is much more computationally demanding than the basic alternative. Thus, even if the extended guess gives more accurate results than the basic guess in a given subspace it may be still beneficial to avoid it and use the basic guess in a larger subspace size instead. For instance, we compare extended-guess 1.0 $N_{\mathrm{MO}}$ calculation for nitroxyl with calculation carried out with basic guess in 2.4 $N_{\mathrm{MO}}$ subspace. It turns out that for either basis set, the basic guess calculations in the larger subspace size are not only more accurate, but also faster. They last approximately 6 and 106 minutes in the aug-cc-pVDZ and aug-cc-pVTZ basis sets, respectively. On the other hand, the extended-guess calculations in the aug-cc-pVDZ basis set last roughly 12 minutes and in the aug-cc-pVTZ basis set -- approximately 131 minutes. This shows that at least in some situations a simple extension of the subspace size might be preferable.

We now go back to the first objection to the practical value of the extended guess, i.e. the fact that in many cases the extended guess gives less accurate results than the basic guess. Despite this shortcoming the extended guess could still be useful if its performance was systematic, i.e. if the obtained errors exhibited some clear dependence on the character of a given state. However, we see that this is not the case. The extended guess works better for singly-excited state of acetylene, but does not show a clear improvement for water molecule. More importantly, it performs very well for some states of genuine higher-excitation character (e.g. nitroxyl calculation in either basis set or nitrosomethane in aug-cc-pVTZ) and fails to provide satisfactory accuracy for other states of the same character (e.g. for glyoxal and cyclobutadiene in either basis set). This behavior cannot be attributed to changes in the compressed triple amplitudes contribution to the vector norm ($r_{XYZ}$) either since, as mentioned previously, these vary only slightly among the investigated genuine higher-excitation systems.

Taking into account the above arguments we cannot recommend the extended guess for practical calculations due to its nonsystematic behavior coupled with the fact that the less expensive alternative in the form of basic guess is available. In a situation when a given subspace size for basic guess calculation does not provide the necessary accuracy, we recommend extending the size of the subspace instead of changing the type of guess. That being said, we do not rule out a possibility that for some systems the extended guess could be better suited and outperform the basic version. Such calculations would however require prior benchmarking on a specific group of systems and thus do not constitute a routine applications which are the focus of the present discussion.   

\subsection{Excited states of magnesium dimer}

We turn our attention to the second part of the calculations in which we investigated the electronic excited states of magnesium dimer. This system has attracted considerable attention in recent years as a promising candidate for studying processes such as coherent control of binary reactions~\cite{rybak11} or collisions at ultracold temperatures~\cite{machholm01}. While $^{24}$Mg$_2$ dimer has numerous advantageous properties, for example, absence of complicated hyperfine structure of energy levels and lack of toxicity unlike its lighter Be$_2$ cousin, the knowledge about properties of the electronic states of this system is far from complete. In this section we show that the rank-reduced EOM-CCSDT is able to fill this gap and determine the potential energy curves and the corresponding spectroscopic parameters for the first four excited singlet states of $^{24}$Mg$_2$.

In the calculations, we adopted aug-cc-pwCVXZ family of basis sets\cite{Prascher2011}, with X = D, T or Q, taken from the Peterson group website.\cite{Peterson_group_website} The triple-excitation subspace size was set to the recommended value of $N_{\mathrm{SVD}}=2N_{\mathrm{MO}}$. This choice is additionally supported by calculation of excitation energies of Mg$_2$ in double- and triple-zeta basis sets in a function of $N_{\mathrm{SVD}}$ for internuclear separations close to the PECs minima (see Supporting Information). These calculations show that $2N_{\mathrm{MO}}$ excitation energies for each of the investigated states are converged to within several thousandth parts of eV with respect to the canonical EOM-CCSDT results. We did not make such preliminary calculations for the quadruple-zeta basis set, however, rank-reduced methods are known to perform better in larger basis sets and thus the above remark safely applies also to the largest basis. Additionally, the eigendecomposition threshold $\sigma_{\mathrm{thr}}$ for these calculations was set to $10^{-6}$. The obtained results were extrapolated to complete basis set (CBS) limit using the following procedure. First, we evaluated the triple- and quadruple-zeta interaction energies for each excited state by subtracting the asymptotic $^1S$\,$+$\,$^1P$ energy of Mg$_2$ in each basis set (calculated at 1000 bohr) from respective potentials. Analogous procedure was applied to the ground state potential where 
the $^1S$\,$+$\,$^1S$ asymptotic energy was subtracted. Next, the interaction energies were extrapolated to CBS limit using the two-point Helgaker extrapolation method\cite{helgaker1997,halkier1998} utilizing the script developed in Ref.~\onlinecite{lang2025}. These results are denoted by the symbol CBS(T,Q) in the remaining discussion. The excited state PECs were subsequently shifted up by the extrapolated excitation energy of the $^1S$\,$+$\,$^1P$ asymptote. For convenience, we set the $^1S$\,$+$\,$^1S$ asymptote to correspond to zero energy in the subsequent analysis.
    
 \begin{figure}[H]
    \centering
    \includegraphics[scale=0.4]{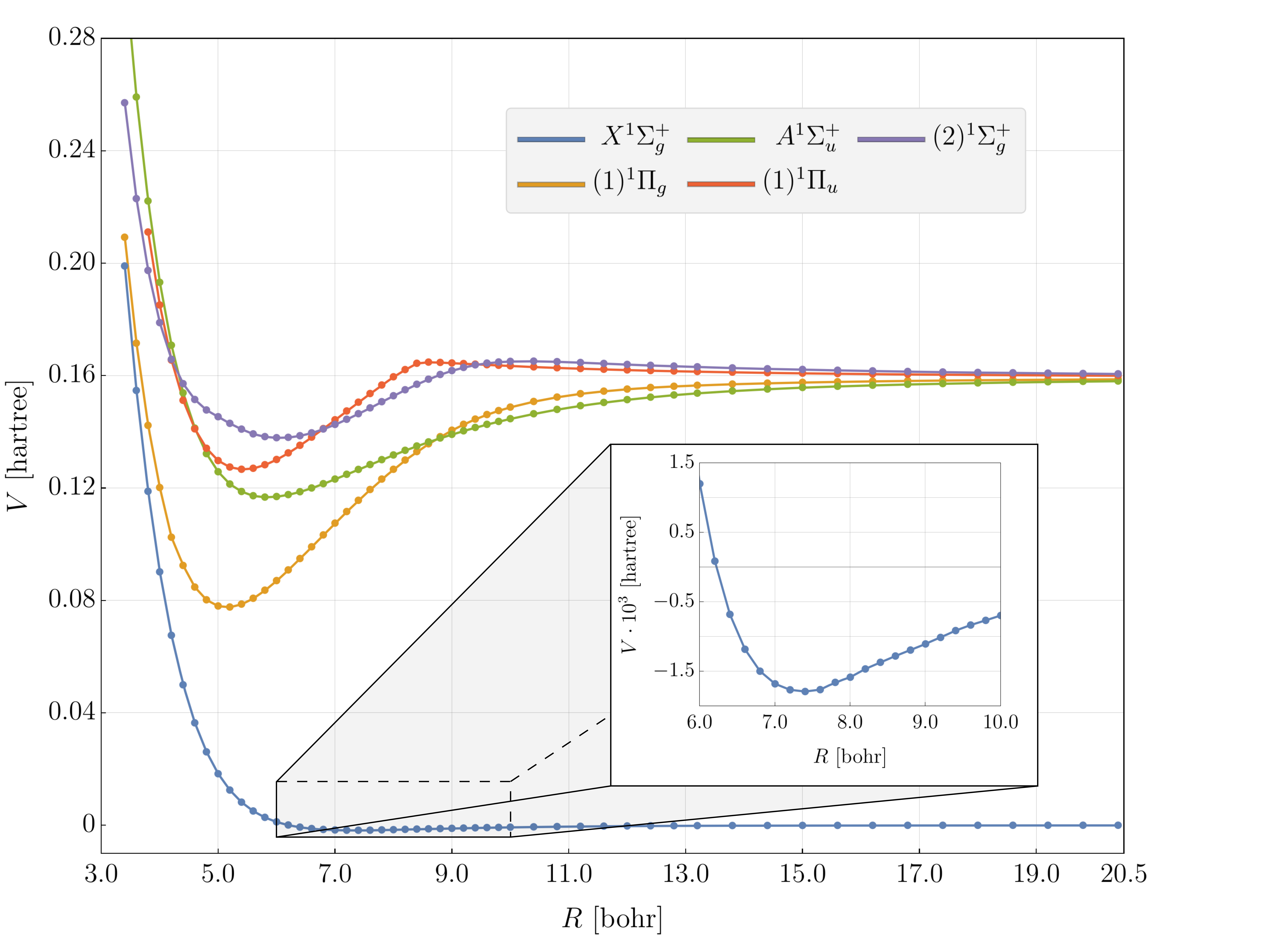}
    \caption{\footnotesize Interaction energy (in hartree) in the ground and the first four singlet excited states of the Mg$_2$ molecule calculated at the RR-EOM-CCSDT/CBS(T,Q) level of theory plotted against the internuclear distance $R$ (in bohr).}
    \label{Mg2_CBS}
\end{figure}

The RR-EOM-CCSDT/CBS(T,Q) interaction energies for the ground state and the first four singlet excited states of Mg$_2$ system are presented in Fig.~\ref{Mg2_CBS} as a function of internuclear distance. The calculations were performed for distances between 3.4 and 10 bohr with 0.2 bohr step size, 10.4-13.2 bohr with 0.4 bohr step size, and 13.8-20.4 bohr with the step size of 0.6 bohr. The $(1)^1\Pi_u$ interaction energy is an exception in this matter as it was determined from the internuclear distance of 3.8 bohr due to convergence difficulties for internuclear distances of 3.4 and 3.6 bohr. However, these points correspond to very high energies in the repulsive wall of the potential and their absence is expected to have tiny impact on the key parameters that characterize the states such as equilibrium distance etc.

A general remark that is appropriate here is that the RR-EOM-CCSDT method produces smooth energy curves free of discontinuities or unphysical bumps. Note that this property is not guaranteed \emph{a priori}, because triple-excitation subspaces are found for each internuclear distance separately. Fortunately, this problem seems either minor or effectively nonexistent and thus the rank-reduced approach can be successfully utilized to generate excited-state potential energy curves.

\begin{table}[H]
\centering
\addtolength{\leftskip} {-2cm}
\addtolength{\rightskip} {-2cm}
 \begin{tabular}{M{1.2cm}|S[table-format=5]|S[table-format=5]|S[table-format=4]|S[table-format=4]|S[table-format=4]|S[table-format=5]|S[table-format=5]|S[table-format=4]}
 \hline
\multirow{2}{*}{State} & 
\multicolumn{1}{M{1.9cm}}{\mbox{\,~~~~~~~~~this work}} & &
\multicolumn{2}{M{2cm}}{\,experiment} & & \multicolumn{2}{M{2cm}}{~~~~theory} \\
\cline{2-9}
 & {AwCQZ} & {CBS(T,Q)} & {Ref.~\onlinecite{Balfour}$^d$} & {Refs~\onlinecite{Knockel_gs, Knockel_exc}}   & {Ref.~\onlinecite{VIDAL197746}$^d$} & {Ref.~\onlinecite{Czuchaj}$^a$}  & {Ref.~\onlinecite{Amaran}$^b$} & {Ref.~\onlinecite{Yuwono}$^c$}
    \\
         \hline
         \,\,$X^1\Sigma_g^+$  & 428
  & 393 & 424 & 430
 & 430  & 404 & 430 & 431
\\
$(1)^1\Pi_g$  & 18019
  & 17953 & {--} & {--}
 & {--}  & 18600  & 18 077 & {--}

\\
\,\,\,$A^1\Sigma_u^+$  & 9401   & 9358   & 9406  & 9414 
 & 9413  & 10480  & 9427  & 9500

\\
$(1)^1\Pi_u$ & 7266
  & 7185 & {--} & 7532  & {--} & 8460  & 5395  & {--}

 \\
$(2)^1\Sigma_g^+$   & 4774
  & 4720 & {--} & {--}
& {--} & 6110 & 2221  & {--}

\\
    \end{tabular}
    \caption{Potential well depths $D_e$ for the ground state and the first four singlet excited states of Mg$_2$. All values have been rounded to the nearest integer and are in units of cm$^{-1}$. Theoretical methods employed in the cited references are briefly summarized in the footnote. }
    \label{De_comparison}
    {\noindent\footnotesize\linespread{1.0}\selectfont \begin{flushleft} 
    $^a$ground state: CCSD(T) (all electrons correlated), excited states: CASSCF/CASPT2, basis set: augmented quintuple-zeta type, counterpoise correction applied. \\
    $^b$ground state: CCSD(T), excited states: LRCCSD, basis set: aug-cc-pVQZ with additional midbond functions, counterpoise correction applied. \\
    $^c$ground state: CCSDT (frozen core) + correction FCI(valence electrons) - CCSDT(valence electrons), excited state: the same scheme but with CR-EOMCCSD(T),IA instead of CCSDT, basis set: aug-cc-pwCVQZ (main calculation), aug-cc-pV(Q + d)Z (correction).  \\
    $^d$the interaction potential depths of the $X^1\Sigma_g^+$ state from Ref.~\onlinecite{VIDAL197746} and the $A^1\Sigma_u^+$ state from Refs.~\onlinecite{Balfour,VIDAL197746} were calculated as the sum of the respective dissociation energies and zero-point vibrational energies. The latter were calculated using the $Y_{i0}$ Dunham coefficients reported in Table V of Ref.~\onlinecite{Balfour} or Table~IV~of \newline Ref.~\onlinecite{VIDAL197746}. 
    \end{flushleft}}
    \end{table}  

Let us now proceed to a quantitative description of the results. All spectroscopic parameters of the calculated potential energy curves discussed in the paper: interaction potential depths $D_e$, dissociation energies $D_0$, equilibrium internuclear distances $R_e$ and harmonic frequencies $\omega_e$ were determined with LEVEL2022 program.\cite{leroy2017} For the computational details regarding the LEVEL2022 calculations see Supporting Information. In order to gauge the accuracy of the RR-EOM-CCSDT method we compare the interaction potential depths $D_e$ for the considered states with the theoretical and experimental results available in the literature, see Table~\ref{De_comparison}.  In the second and third column of Table~\ref{De_comparison} we list the results of our calculations obtained in the aug-cc-pwCVQZ basis set (denoted shortly as AwCQZ) and in the CBS(T,Q) limit, respectively. In the remaining columns of Table~\ref{De_comparison} we gathered both experimental and theoretical values of $D_e$ reported in the literature. We stress that this list is not exhaustive and focuses on the most recent papers which include results for the excited states under consideration.

Let us start the discussion with comparison of the ground-state calculations. The obtained AwCQZ and CBS(T,Q) curves for $X^1\Sigma_g^+$ state are characterized by $D_e$ of 428 and 393 cm$^{-1}$, respectively. The AwCQZ result is in good agreement with experimental and theoretical values with deviations of the order of a few cm$^{-1}$. The only exception is the theoretical result obtained by Czuchaj et al. which is smaller by 26 cm$^{-1}$ with respect to the value of 430 cm$^{-1}$ that was obtained in the most recent experiments. Our CBS(T,Q) result differs from the aforementioned experimentally-derived value by 37 cm$^{-1}$. The fact that the extrapolated result shows larger disagreement with the experimental result than the AwCQZ data is troubling. First, we suspected that the extrapolation method adopted in this work performs unusually poorly in this case. However, we tested a different extrapolation protocol\cite{lesiuk2019b} based on the Riemann zeta function and obtained very similar results. Therefore, we suspect that the agreement between the results obtained with the AwCQZ basis set and the experimental data may be just a consequence of an accidental (partial) cancellation of errors between the basis set incompleteness and other sources of error. Fortunately, this phenomenon is much less problematic for excited states of Mg$_2$ where the AwCQZ and CBS(T,Q) results differ by a much smaller margin (on a relative basis). To gauge this effect, we continue to provide AwCQZ and CBS(T,Q) results separately for all data considered.

Next, we turn our attention to the $A^1\Sigma_u^+$ and $(1)^1\Pi_u$ excited states for which the experimental data is also available. The three experimentally-derived $A^1\Sigma_u^+$ potential depths agree well with each other, as the largest discrepancy reaches only 8 cm$^{-1}$, see Table~\ref{De_comparison}. The theoretical results obtained with the LRCCSD method by Amaran et al. are only 13 cm$^{-1}$ larger than the most recent experimental value reported by Kn\"ockel and coworkers.  The composite CR-EOMCCSD(T),IA/FCI result by Yuwono et al. differs by 86 cm$^{-1}$ from this value, and a significantly larger discrepancy is observed in the case of the CASSCF/CASPT2 calculations. We observe that our AwCQZ result for $A^1\Sigma_u^+$ state is only 13 cm$^{-1}$ smaller than the most recent experimental value, while the discrepancy in the CBS(T,Q) limit is roughly four times larger. Therefore, the AwCQZ result again agrees better with the experimental data, but the relative difference between AwCQZ and CBS(T,Q) values is drastically smaller than for the ground state (ca. 8\% vs. 0.5\%). The $D_e$ parameters obtained in this work for the $(1)^1\Pi_u$ state in the AwCQZ basis set and in the CBS(T,Q) limit differ from the only available experimentally-derived result by 266 and 347 cm$^{-1}$, respectively. These errors are significantly larger than the errors encountered for the $A^1\Sigma_u^+$ state. However, we note that other theoretical results reported for this state in the literature also disagree with the experimental result, see Table~\ref{De_comparison}. Our results sit in between theoretical values reported in Refs.~\onlinecite{Czuchaj,Amaran} and have considerably smaller errors with respect to the experimental data. 

In the case of the excited $(2)^1\Sigma_g^+$ state, where no experimental data is available, the situation is quite similar as in the $A^1\Sigma_u^+$ state in the sense that the $D_e$ parameters obtained in AwCQZ and CBS(T,Q) calculations are in between the theoretical values reported in Refs.~\onlinecite{Czuchaj,Amaran}, which differ by nearly 3900 cm$^{-1}$. The potential depths obtained in this work are smaller than the CASSCF/CASPT2 values by roughly 1300-1400 cm$^{-1}$ and larger than the LRCCSD values by approximately 2500-2550 cm$^{-1}$. The large discrepancy with respect to the latter result may be (at least partially) explained by the inclusion of full triple-excitation amplitudes in the coupled cluster method utilized in the present work. Lastly, we turn our attention to the results obtained for the $(1)^1\Pi_g$ state which also has not been observed experimentally. Our AwCQZ and CBS(T,Q) results differ from the LRCCSD value by 58 and 124 cm$^{-1}$, respectively, while the discrepancies between them and the CASSCF/CASPT2 calculations equal 581 and 647 cm$^{-1}$. Although the theoretical values differ quite significantly, these differences are much smaller than for the $(2)^1\Sigma_g^+$ state.

Next, we discuss other spectroscopic parameters calculated from the RR-EOM-CCSDT/RR-CCSDT interaction potential curves ($D_0$, $R_e$, $\omega_e$) for the $^{24}$Mg$_2$ molecule. These parameters are gathered for all investigated states in Table~\ref{Spectroscopic_constants}, which, for convenience, includes also the potential depths $D_e$ discussed above. 

In general, the values of $R_e$ presented in Table~\ref{Spectroscopic_constants} agree reasonably well with previous experimental and theoretical studies. The AwCQZ results differ from the most recent experimentally-derived values of Kn\"ockel et al. for the states $X^1\Sigma_g^+$, $A^1\Sigma_u^+$, $(1)^1\Pi_u$ by roughly 0.043, 0.003 and 0.014 \AA, respectively. The discrepancies with respect to other theoretical data are generally of the order of 0.01-0.05 \AA. A notably larger difference of approximately 0.1 \AA~ is encountered for $(2)^1\Sigma_g^+$ state in comparison with the result reported by Amaran et al. In contrast to the interaction energies, the errors in CBS(T,Q) $R_e$ parameters are only slightly different from their AwCQZ counterparts as the $R_e$ values themselves calculated with both schemes are very close to each other - the largest difference being 0.009 \AA~ for $(2)^1\Sigma_g^+$ state. The calculated $\omega_e$ values are also quite satisfactory.  Comparing our AwCQZ values against the experimental results by Kn\"ockel and coworkers, the errors for states $X^1\Sigma_g^+$, $A^1\Sigma_u^+$, $(1)^1\Pi_u$ equal roughly 2, 5 and 2 cm$^{-1}$. The errors for the last two states are slightly larger for $\omega_e$ values obtained in CBS(T,Q) limit and equal approximately 6 and 5 cm$^{-1}$. The errors with respect to the values obtained by Czuchaj et al. are of similar magnitude, the largest being roughly 8 cm$^{-1}$ for  $(2)^1\Sigma_g^+$ state (as compared with our AwCQZ result). 

\begin{table}[H]
\centering
 \begin{tabular}{M{1.2cm}|S[table-format=5]|S[table-format=5]|S[table-format=1.3]|S[table-format=3]|S[table-format=5]|S[table-format=5]|S[table-format=1.3]|S[table-format=3]}
 \hline
 \multirow{2}{*}{State} &
 \multicolumn{3}{M{2cm}}{AwCQZ} && \multicolumn{3}{M{2cm}}{CBS(T,Q)} \\
 \cline{2-9}
    & {$D_e$} & {$D_0$} & {$R_e$} & {$\omega_e$}  & {$D_e$} & {$D_0$} & {$R_e$} & {$\omega_e$}   \\
         \hline
         \,\,$X^1\Sigma_g^+$  & 428 & 402
  & 3.933
  & 49    & 393 & 365
 & 3.933
& 49
\\
$(1)^1\Pi_g$  & 18019 & 17874
  & 2.721
  &  290 & 17953 & 17809
 & 2.722
 & 289
\\
\,\,\,$A^1\Sigma_u^+$  & 9401 & 9307
   &  3.085
     & 186  & 9358 & 9264
  & 3.085
 & 185
\\
$(1)^1\Pi_u$ & 7266 & 7142
  & 2.870
    & 248  & 7185 & 7062
 & 2.875
& 245
 \\
$(2)^1\Sigma_g^+$   & 4774 & 4690
  & 3.191    & 170 & 4720 & 4637 & 3.200
 & 166
\\
    \end{tabular}
    \caption{Spectroscopic parameters for the ground state and the first four singlet excited states of Mg$_2$. Values of $D_e$, $D_0$ and $\omega_e$ are in cm$^{-1}$, while $R_e$ is given in angstrom.}
    \label{Spectroscopic_constants}
    \end{table}

\subsection{Charge-transfer excitation}

Charge-transfer excitations are encountered in many different research areas, such as in photovoltaics or photocatalysis, and in many cases their properties are crucial for understanding of the underlying photochemical processes.\cite{coropceanu2019,may2024}
Simultaneously, these types of excitations are difficult to describe accurately with TDDFT\cite{sobolewski2003,dreuw2003,dreuw2004,mester2022} or even more sophisticated and expensive methods such as EOM-CCSD\cite{kozma2020,izsak2020}.
The reason for this troublesome behavior appears to be, at least to a significant degree, the consequence of substantial orbital relaxation effects not limited to linear terms with respect to the reference determinant.\cite{subotnik11,herbert2023}
Simultaneously, charge-transfer excitations may occur over distances of tens, if not hundreds, of angstroms which makes techniques based on arguments of orbital localization problematic. As a result, it is interesting to verify whether rank-reduced approaches can be applied to such long-range charge-transfer excitations and whether, for example, stable levels of accuracy can be achieved as a function of distance between donor and acceptor.

We tested the RR-EOM-CCSDT method in calculation of the potential energy curve corresponding to the intermolecular charge-transfer excitation in the $\mathrm{NH}_3$-$\mathrm{F}_2$ system. The charge-transfer excitation energy was determined for distances between the monomers in the range of 6-100 bohr, measured between the center of the fluorine molecule and the nitrogen atom of ammonia. This distance will be denoted by the symbol $R$ further in the text. We fully optimized the geometry of the system for smaller $R$, namely between 6-15 bohr with step size of 0.5 bohr and for 20 bohr, while the geometries for distances greater than 20 bohr were constructed from the optimized geometry for 20 bohr by moving the $\mathrm{NH}_3$ molecule further away from the fluorine in a direction parallel to the molecular axis of F$_2$. The step size for the latter procedure was 5 bohr for the distances up to 50 bohr, and 10 bohr for larger $R$ up to 100 bohr. The geometry optimization for smaller distances was carried out at the B3LYP-D3\cite{becke1993,stephens1994,grimme2010}/aug-cc-pVTZ level of theory using the Gaussian~16 program\cite{g16}. The starting points for the geometry optimization were obtained from Ref.~\onlinecite{kozma2020} by moving the ammonia molecule to a desired distance away from $\mathrm{F}_2$ (as described above). The optimization was then carried out using the AddGIC keyword in the Gaussian program which allowed to freeze the distance between the center of the fluorine molecule and the nitrogen atom of ammonia, while allowing for  relaxation of the remaining degrees of freedom. Thresholds for convergence of this procedure were lowered with the Tight keyword and the SuperFine grid for DFT integration was used. In the excited-state calculations for the obtained geometries we used Dunning-type cc-pVDZ and cc-pVTZ basis sets.\cite{dunning1989a} The RR-EOM-CCSDT calculations were carried out with both basic and extended types of guess and for triple-excitation subspace sizes $N_{\mathrm{SVD}}$ equal to 1~$N_{\mathrm{MO}}$ , 1.5~$N_{\mathrm{MO}}$, 2~$N_{\mathrm{MO}}$, 2.5~$N_{\mathrm{MO}}$ and 3~$N_{\mathrm{MO}}$. Lastly, in the HOOI procedure for this system we used the exact, diagonalized $T_{ij}^{ab}$ and $R_{ij}^{ab}$ amplitudes, i.e. we retained all the eigenvalues in the decomposition.

\begin{figure}[H]
    \centering
    \includegraphics[scale=0.6]{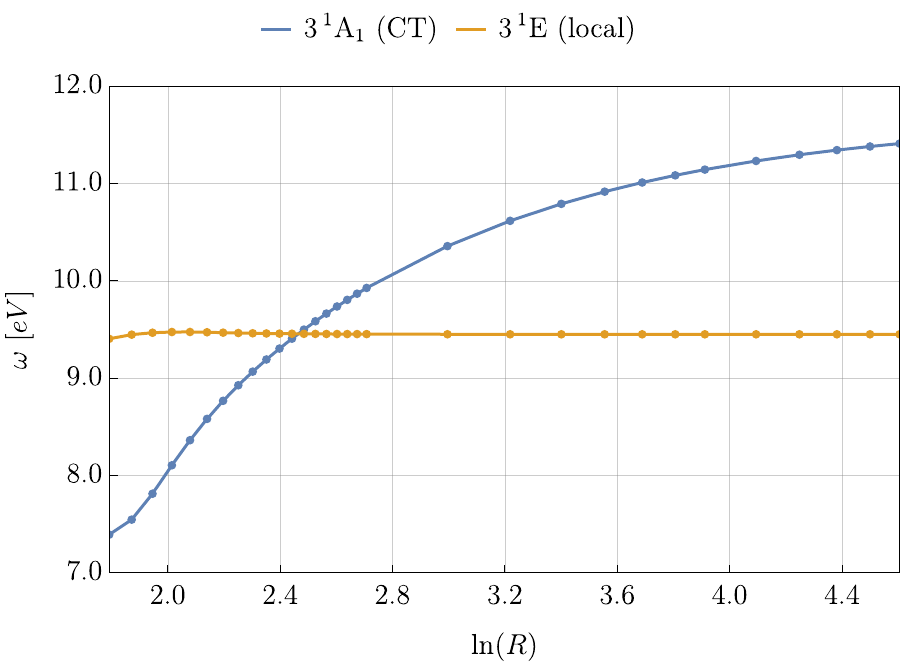}
    \caption{\footnotesize RR-EOM-CCSDT/cc-pVTZ excitation energies of the $3\,{^1\mathrm{A}}_1$ charge-transfer state (blue curve) and the $3\,^1\mathrm{E}$ state (orange curve) of the $\mathrm{NH}_3$-$\mathrm{F}_2$ system plotted against the distance between the monomers $R$ (logarithmic scale on the horizontal axis). For large $R$ the excitation in the $3\,^1\mathrm{E}$ state is localized on the ammonia molecule.}
    \label{lnNH3-F2_3nmo_ext_plot}
\end{figure}

We start the discussion with some general information about the investigated charge-transfer excitation following Ref.~\onlinecite{kozma2020b}. For illustrative purposes, in Fig.~\ref{lnNH3-F2_3nmo_ext_plot} we present the RR-EOM-CCSDT/cc-pVTZ excitation energy curves calculated with the extended guess and $N_{\mathrm{SVD}}$ = 3~$N_{\mathrm{MO}}$ as an example. The distance between the monomers $R$ ranges from 6 up to 100 bohr and is shown on a logarithmic scale. The charge-transfer (CT) curve is marked with blue color and corresponds to the $3\,^1\mathrm{A}_1$ electronic state (the complex retains the $C_{3v}$ symmetry for all of the investigated distances). The CT excitation energy shows strong dependence on the intermolecular distance and for large $R$ we observe $1/R$ behavior consistent with Coulomb interaction of two charges. The difference in the calculated excitation energy between $R=6$ bohr and $R=100$ bohr is roughly 4 eV. The CT curve crosses the 3$\,^1\mathrm{E}$ state (marked with orange) at approximately 11.77 bohr. The 3$\,^1\mathrm{E}$ state corresponds to an excitation localized on ammonia and thus, in this case, the excitation energy does not show a significant dependence on $R$ -- it ranges between roughly 9.40 and 9.47 eV. Lastly, we note that the curves obtained in the rank-reduced approach are smooth and do not show any non-analytic or otherwise unphysical behavior. This is achieved despite the strong distance dependence of the CT excitation energy on $R$, indicating that the decomposition errors remain relatively small with varying monomer separation. 

Next, we turn our attention to the description of the decomposition errors in a quantitative manner. To this end, we compare the RR-EOM-CCSDT/cc-pVDZ excitation energies for the aforementioned intermonomer distances with the reference values. The reference energies are obtained with the RR-EOM-CCSDT method in the complete excitation subspace ($N_\mathrm{SVD} = OV$) which is equivalent to the canonical EOM-CCSDT theory. To make sure that the RR-EOM-CCSDT program has no implementation error that would spoil the accuracy even in the $N_\mathrm{SVD} = OV$ limit, we additionally compared our results with the canonical EOM-CCSDT results for distances 13.5, 45, 70 and 100 bohr that were obtained using the Q-Chem program package.\cite{qchem} The thresholds for convergence of the Davidson's procedure and for including a new expansion vector in the iterations were set to $10^{-7}$ in the Q-Chem calculations. All other input parameters were kept at their default values. The results obtained by two programs differed by no more than 0.0001-0.0002 eV which ensures that the RR-EOM-CCSDT excitation energies calculated in the full space indeed converge to the exact EOM-CCSDT values and can be used as a reference.

The absolute errors in the excitation energy of the CT state are presented as a function of monomer separation $R$ in Figs.~\ref{lnNH3-F2_bg_dz_exc_en}~and~\ref{lnNH3-F2_extg_dz_exc_en}, which correspond to the calculations with the basic and extended types of guess, respectively. The results are presented for triple excitation subspace sizes between 1.0 and 3.0 times $N_{\mathrm{MO}}$. We note that in both types of guess a tiny error for $R\approx6-7$ bohr is artificial as around these points the error crosses zero. Focusing on Fig.~\ref{lnNH3-F2_bg_dz_exc_en}, we observe that the errors behave similarly in a function of $R$ for all subspace sizes considered. The errors rise from the initially small (artificial) values, then plateau for $R$ equal to approximately 15 bohr and show only minor fluctuations as $R$ is increased further. The only exception from this behavior is exhibited by a single point in 1 $N_\mathrm{MO}$ plot  which shows a much smaller error than expected (at $R=35$ bohr). 

We observe a clear convergence toward the exact result with the growing size of the subspace. As expected, the largest errors are encountered in 1$N_\mathrm{MO}$ and 1.5$N_\mathrm{MO}$ subspace sizes. For the 2$N_\mathrm{MO}$ subspace we see an approximately twofold reduction in error in comparison to the smaller subspaces. Extending the subspace size further still decreases the error (although less dramatically) to values around 0.030-0.034 eV for 3$N_\mathrm{MO}$ subspace. We note that this error is still significant, considering the dominant singly-excited character of the charge-transfer excitation. On the other hand, it is roughly a hundred times smaller than the aforementioned 4 eV difference between the CT excitation energies calculated at 6 and 100 bohr.  

We now turn our attention to Fig.~{\ref{lnNH3-F2_extg_dz_exc_en} which shows the results obtained with the extended guess. We notice that the extended-guess errors for $R \ge$ 7.50 bohr are almost universally smaller than the basic-guess errors, irrespective of the size of the subspace. The only exception consists of 1$N_\mathrm{MO}$ errors which rise up to around 0.065 eV for larger distances. These errors are however still smaller than their basic-guess counterparts in the 1$N_\mathrm{MO}$ subspace. We notice that in the case of extended guess, the errors are not decreasing monotonically as the subspace is enlarged, in contrast with the basic guess results. Additionally, the error obtained for the largest subspace size with the extended guess is roughly four times smaller than the corresponding results with the basic guess. This makes the extended guess preferable for this particular problem, although, in practice, going to the larger subspace sizes with the basic guess might still prove to be more economical. 

 \begin{figure}[H]
    \centering
    \includegraphics[scale=0.6]{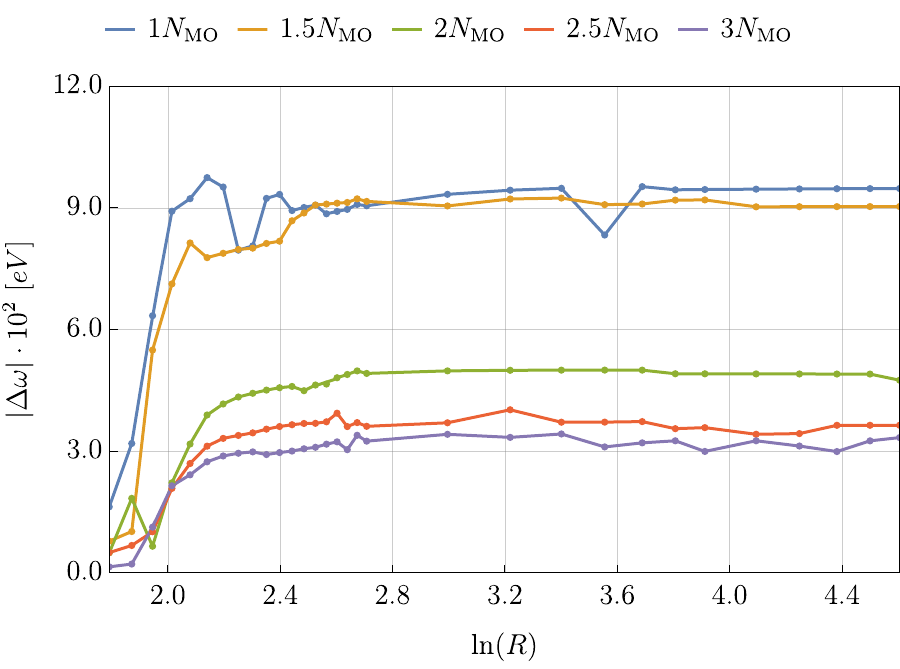}
    \caption{\footnotesize Absolute errors in the excitation energy of the charge-transfer state in the $\mathrm{NH}_3$-$\mathrm{F}_2$ system calculated at the RR-EOM-CCSDT/cc-pVDZ level of theory (basic guess) plotted against the distance between the monomers $R$ (logarithmic scale on the horizontal axis). The calculations were made for five dimensions of the triple-excitation subspace, see the legend. The error was measured against the RR-EOM-CCSDT/cc-pVDZ calculation carried out in the complete triple-excitation subspace which is equivalent to the canonical EOM-CCSDT method.}
    \label{lnNH3-F2_bg_dz_exc_en}
\end{figure}

 \begin{figure}[H]
    \centering
    \includegraphics[scale=0.6]{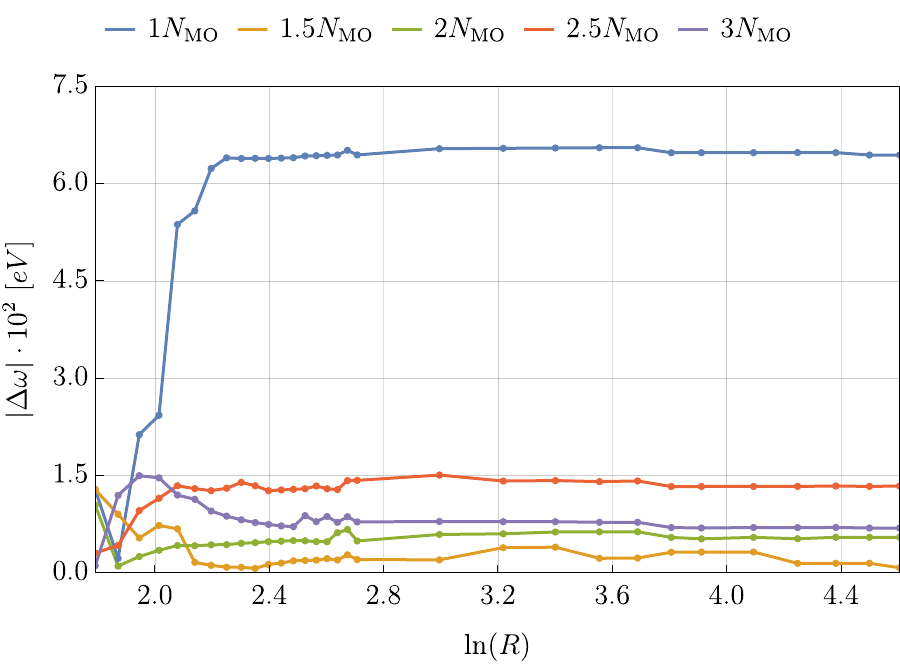}
    \caption{\footnotesize 
    Absolute errors in the excitation energy of the charge-transfer state in the $\mathrm{NH}_3$-$\mathrm{F}_2$ system calculated at the RR-EOM-CCSDT/cc-pVDZ level of theory (extended guess) plotted against the distance between the monomers $R$ (logarithmic scale on the horizontal axis). The calculations were made for five dimensions of the triple-excitation subspace, see the legend. The error was measured against the RR-EOM-CCSDT/cc-pVDZ calculation carried out in the complete triple-excitation subspace which is equivalent to the canonical EOM-CCSDT method.}
    \label{lnNH3-F2_extg_dz_exc_en}
\end{figure}

We now move to the discussion of the cc-pVTZ results which are presented for basic and extended guess in Figs.~\ref{lnNH3-F2_bg_tz_exc_en_nmo_diff}~and~\ref{lnNH3-F2_extg_tz_exc_en_nmo_diff}, respectively. Due to the lack of a proper reference values we focus in this case only on the convergence of our results with respect to the excitation subspace size. As above, we consider five dimensions of this subspace: 1$N_\mathrm{MO}$, 1.5$N_\mathrm{MO}$, 2$N_\mathrm{MO}$, 2.5$N_\mathrm{MO}$ and 3$N_\mathrm{MO}$. In Figs.~\ref{lnNH3-F2_bg_tz_exc_en_nmo_diff}~and~\ref{lnNH3-F2_extg_tz_exc_en_nmo_diff} we plot absolute differences between the excitation energies obtained with each two consecutive subspace sizes from the above set (as a function of the intermolecular distance $R$). For example, the results corresponding to the difference between the  1.5$N_\mathrm{MO}$ and  1$N_\mathrm{MO}$ excitation energies are denoted as $\Delta \omega_{1.5-1.0}$. For both types of guess we achieve the convergence to the exact result to within 0.01 eV, although at different rates. In the case of the basic guess we notice that the $|\Delta \omega_{1.5-1.0}|$ values stabilize at around 0.049-0.052 eV for larger distances. The difference between the results obtained with the next pair of subspace sizes is roughly 3 times smaller and  stabilizes at approximately 0.016-0.018 eV. Lastly, the values obtained for the largest subspaces ($|\Delta \omega_{3.0-2.5}|$ and $|\Delta \omega_{2.5-2.0}|$ ) are very similar to each other and both lie significantly below 0.01 eV, indicating the convergence with respect to the subspace size has been achieved. The differences $|\Delta \omega_{1.5-1.0}|$ calculated with the extended guess are noticeably smaller than in the case of the basic guess, with the values between 0.0216-0.0235 eV for larger intermonomer distances. Moreover, we observe that the calculated differences drop to values below 0.01 eV when larger subspaces are considered. Lastly, it is worth mentioning that the $|\Delta \omega_{3.0-2.5}|$ values lie between the $|\Delta \omega_{2.5-2.0}|$ and $|\Delta \omega_{2.0-1.5}|$ results for $R \ge 7.5$ bohr, indicating that the convergence is not monotonic again. Nevertheless, all of these differences lie below 0.01 eV and deviations between them are small enough to expect that any further extension of the subspace does not change the results significantly.

 \begin{figure}[H]
    \centering
    \includegraphics[scale=0.6]{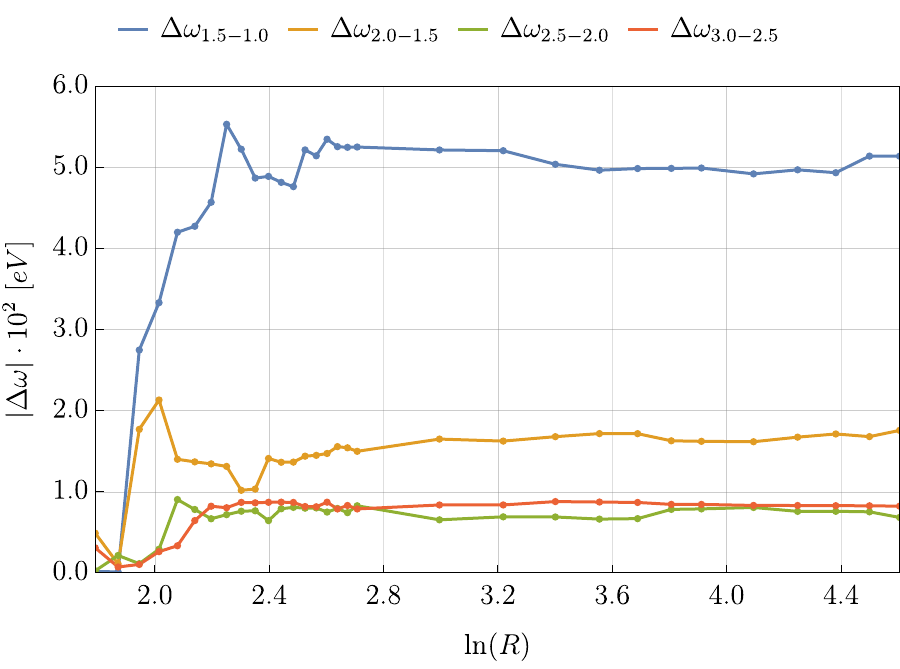}
    \caption{\footnotesize Absolute differences in the excitation energies calculated with pairs of two consecutive triple-excitation subspace sizes for the charge-transfer state of the $\mathrm{NH}_3$-$\mathrm{F}_2$ system (RR-EOM-CCSDT/cc-pVTZ level of theory, basic guess). The plot corresponding to the difference between the $x \cdot N_\mathrm{MO}$ and $y \cdot N_\mathrm{MO}$ results is denoted by the symbol $\Delta \omega_{x-y}$, see the legend.}
    \label{lnNH3-F2_bg_tz_exc_en_nmo_diff}
\end{figure}

 \begin{figure}[H]
    \centering
    \includegraphics[scale=0.6]{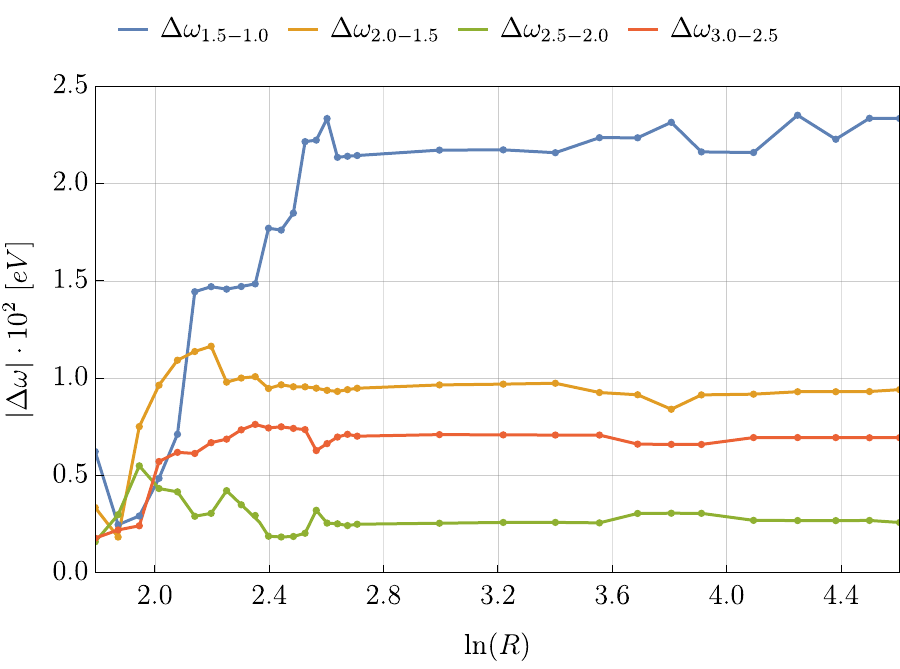}
    \caption{\footnotesize Absolute differences in the excitation energies calculated with pairs of two consecutive triple-excitation subspace sizes for the charge-transfer state of the $\mathrm{NH}_3$-$\mathrm{F}_2$ system (RR-EOM-CCSDT/cc-pVTZ level of theory, extended guess). The plot corresponding to the difference between the $x \cdot N_\mathrm{MO}$ and $y \cdot N_\mathrm{MO}$ results is denoted by the symbol $\Delta \omega_{x-y}$, see the legend.}
    \label{lnNH3-F2_extg_tz_exc_en_nmo_diff}
\end{figure}

\section{Conclusions and Outlook}
\label{sec:outlook}

In this paper, we extend the formalism described in our previous work on the RR-EOM-CC3 method to fully account for triple excitations. The proposed scheme, termed RR-EOM-CCSDT, has the formal $N^6$ scaling with the system size in contrast with the $N^8$ scaling of its parent EOM-CCSDT method. It relies on solving the EOM-CCSDT eigenvalue problem where the ground- and excited-state triple-excitation amplitudes are represented in the Tucker format and the triple amplitudes equation is projected onto the excited-state triple-excitation subspace obtained from the decomposition of approximate $R_{ijk}^{abc}$ amplitudes. These approximate amplitudes are determined through the perturbative expansion of the EOM-CCSDT triple amplitudes equation. We introduced basic and extended versions of guess based on this procedure, where basic guess is correct to the first order in the expansion while extended guess includes two selected second-order terms as well. The key parameter of the developed method is the triple-excitation subspace size $N_{\mathrm{SVD}}$ which is usually set equal for both the ground- and excited-state subspaces. The choice of subspace size is a trade-off between the accuracy and affordability of calculations with the full space $N_{\mathrm{SVD}}=OV$ recovering the exact EOM-CCSDT result. 

The accuracy of the RR-EOM-CCSDT method was assessed in a series of calculations. In the first part, we focused on isolated excited states of ten molecules which exhibit varied character -- from nearly pure single- to dominated by higher-excitations. Comparison of the RR-EOM-CCSDT results with the exact EOM-CCSDT excitation energies shows that the errors several times smaller than the inherent error of EOM-CCSDT for a given state are achievable. We recommend 2$N_{\mathrm{MO}}$ subspace size for both aug-cc-pVDZ and aug-cc-pVTZ basis sets as the errors in this case do not exceed 0.03 eV for all considered systems. The obtained data was also used to assess the viability of basic and extended types of guess. On average, the accuracy of both approaches is comparable for the investigated systems, but the results differ widely in some individual cases. We recommend the basic guess over the extended version due to the non-systematic performance of the latter approach and its much higher cost.

We also evaluated the suitability of the RR-EOM-CCSDT method in calculation of excited-state PECs of diatomic molecules and charge-transfer systems on the examples of Mg$_2$ and NH$_3$-F$_2$. In both cases we obtain smooth curves devoid of any unphysical behaviour, despite the necessity of determining the triple-excitation subspaces separately for each individual distance. This is especially noteworthy for the NH$_3$-F$_2$ system, where the excitation energy of the charge-transfer state changes significantly with the intermonomer separation. Moreover, the calculations for Mg$_2$ show that the RR-EOM-CCSDT method is not only qualitatively correct as the spectroscopic parameters obtained for excited states of this system agree well with available  experimental data.

In conclusion, the RR-EOM-CCSDT method proposed in this work is able to provide excitation energies of near-EOM-CCSDT quality at significantly reduced computational cost and more manageable memory requirements. These facts make the presented method a promising candidate for accurate calculations of excitation energies of large molecules, for which EOM-CCSDT is prohibitively expensive. It would also be interesting to see how well the proposed method performs as a correction in various composite schemes, replacing its canonical equivalent. Lastly, the theory described in the present work is a necessary foundation before going into quadruple excitations with the rank-reduced approach. In particular, the rank-reduced EOM-CC4 method would be a highly desirable extension.

\begin{acknowledgement}
The work was supported by the National Science Centre, Poland, under research project 2022/47/D/ST4/01834. We gratefully acknowledge Poland's high-performance Infrastructure PLGrid (HPC Centers: ACK Cyfronet AGH, PCSS, CI TASK, WCSS) for providing computer facilities and support within computational grant PLG/2023/016599.
\end{acknowledgement}

\begin{suppinfo}
The following files are available free of charge:
\begin{itemize}
  \item {\tt Supporting\_Information.pdf}: details of calculations in LEVEL2022 program, additional information about convergence of Mg$_2$ excitation energies with the subspace size and molecular structures of investigated systems. 
  \item {\tt RR-EOM-CCSDT\_derivation.pdf}: derivation of the RR-EOM-CCSDT equations, explicit formula for the $\Omega_{XYZ}$ residual tensor.
  \item {\tt Guess\_derivation.pdf}: derivation of the triple amplitudes guess equations, explicit formula for the projected $R_{ia,YZ}$ amplitude.
\end{itemize}

\end{suppinfo}

\bibliography{rr-eom-ccsdt-paper}
\end{document}